\documentclass[12pt]{amsart}

\usepackage{ljm-auth}
\usepackage{graphicx}
\usepackage{pgf}

\usepackage[normalem]{ulem}

\newcommand{\comment}[1]{}
\newcommand{\quotes}[1]{``#1''}

\newcommand{\cnot}{\mathrm{CNOT}}

\newcommand{\PHASE}[1]{\mathrm{PHASE\left(#1\right)}}

\newcommand{\mat}{\left( \!\! \begin{array}{cc}}
\newcommand{\rix}{\end{array} \!\! \right)}

\usepackage{color}
\def\bR{\begin{color}{red}}
\def\bB{\begin{color}{blue}}
\def\bM{\begin{color}{magenta}}
\def\bC{\begin{color}{cyan}}
\def\bW{\begin{color}{white}}
\def\bBl{\begin{color}{black}}
\def\bG{\begin{color}{green}}
\def\bY{\begin{color}{yellow}}
\def\ec{\end{color}\ }

\usepackage{xy}
\xyoption{matrix}
\xyoption{frame}
\xyoption{arrow}
\xyoption{arc}

\usepackage{ifpdf}
\ifpdf
\else
\PackageWarningNoLine{Qcircuit}{Qcircuit is loading in Postscript mode.  The Xy-pic options ps and dvips will be loaded.  If you wish to use other Postscript drivers for Xy-pic, you must modify the code in Qcircuit.tex}
\xyoption{ps}
\xyoption{dvips}
\fi

\entrymodifiers={!C\entrybox}

\newcommand{\ket}[1]{{\left\vert{#1}\right\rangle}}
\newcommand{\qw}[1][-1]{\ar @{-} [0,#1]}
\newcommand{\qwx}[1][-1]{\ar @{-} [#1,0]}

\newcommand{\gate}[1]{*+<.6em>{#1} \POS ="i","i"+UR;"i"+UL **\dir{-};"i"+DL **\dir{-};"i"+DR **\dir{-};"i"+UR **\dir{-},"i" \qw}

\newcommand{\control}{*!<0em,.025em>-=-<.2em>{\bullet}}

\newcommand{\ctrl}[1]{\control \qwx[#1] \qw}

\newcommand{\targ}{*+<.02em,.02em>{\xy ="i","i"-<.39em,0em>;"i"+<.39em,0em> **\dir{-}, "i"-<0em,.39em>;"i"+<0em,.39em> **\dir{-},"i"*\xycircle<.4em>{} \endxy} \qw}
\newcommand{\qswap}{*=<0em>{\times} \qw}
\newcommand{\multigate}[2]{*+<1em,.9em>{\hphantom{#2}} \POS [0,0]="i",[0,0].[#1,0]="e",!C *{#2},"e"+UR;"e"+UL **\dir{-};"e"+DL **\dir{-};"e"+DR **\dir{-};"e"+UR **\dir{-},"i" \qw}
\newcommand{\ghost}[1]{*+<1em,.9em>{\hphantom{#1}} \qw}

\newcommand{\lstick}[1]{*!R!<.5em,0em>=<0em>{#1}}

\newcommand{\Qcircuit}{\xymatrix @*=<0em>}

\newcommand{\ET}[1]{\mathrm{ET\left(#1\right)}}
\newcommand{\CET}{\mathrm{C(ET)}}

\begin{document}

\bibliographystyle{unsrt}

\author{F.\,M.\,Ablayev, S.\,N.\,Andrianov, S.\,A.\,Moiseev, A.\,V.\,Vasiliev}
\crauthor{Ablayev, Andrianov, Moiseev, Vasiliev} 

\tit{Quantum computer with atomic logical qubits encoded on macroscopic
three-level systems in common quantum electrodynamic cavity}
\shorttit{Quantum computer in common QED cavity} 

\setcounter{page}{3}

\maketit

\begin{center}{Kazan Federal University\\%
Kazan Physical-Technical Institute of the Russian Academy of
Sciences\\%
Institute for Informatics of Tatarstan Republic Academy of
Sciences\\%
}
\end{center}
\email{fablayev@gmail.com, andrianovsn@mail.ru, samoi@yandex.ru,
alexander.ksu@gmail.com}

\abstract{We propose an effective realization of the  universal set of
elementary quantum gates in solid state quantum computer based on macroscopic
(or mesoscopic) resonance systems -- multi-atomic coherent ensembles, squids or
quantum dots in quantum electrodynamic cavity. We exploit an encoding of
logical qubits by the pairs of the macroscopic two- or three-level atoms that
is working in a Hilbert subspace of all states inherent to these atomic
systems. In this subspace, logical single qubit gates are realized by the
controlled reversible transfer
of single atomic excitation in the pair via the exchange of virtual photons and
by the frequency shift of one of the atomic ensembles
in a pair. In the case of two-level systems, the logical two-qubit gates are
performed by the controlling of Lamb shift magnitude in one atomic ensemble,
allowing/blocking the excitation transfer in a pair, respectively, that is
controlled by the third atomic system of another pair. When using three-level
systems, we describe the NOT-gate in the atomic pair controlled by the transfer
of working atomic excitation to the additional third level caused by direct
impact of the control pair excitation. Finally, we discuss advantages of the
proposed physical system for accelerated computation of some useful quantum
gates.}

\notes{0}{%
\thank{Partially supported by Russian Foundation for Basic Research, Grants
11-07-00465, 12-01-31216, 12-07-97016\_p}}


\section{Introduction}

Various physical systems using single natural or artificial atoms, ions,
molecules etc. have been proposed during two last decades for the construction
of a quantum computer
\cite{Nielsen-Chuang:2000:QC,Nakahara-Ohmi:2008:QC,Kok:2007:OpticalQC,Ladd:2010:QC}.
Creation of a quantum computer with a large number of qubits is a huge problem
on the known systems, first of all, because of too strong decoherence of the
qubits. That makes the search for the new physical systems and relevant
experimental approaches still actual. One of the promising approaches is using
natural atoms (ions, molecules, ...) with long coherence time.

Recently, new physical realization of a quantum computer based on the
\textit{multi-atomic coherent} (MaC) ensembles has been proposed for encoding
of separate qubits \cite{Brion:2007:Ensembles,Saffman-Molmer:2008:RydbergQC}.
MaC ensembles yield huge amplification of dipole moment on the resonance
transition that leads to essential acceleration of the quantum information
processing rate. However, transitions to the excess states in a MaC ensemble
should be blocked in order to realize effective two-level systems providing an
ideal encoding of qubits. Dipole-dipole interaction is intensively discussed
for the blockade of excess quantum states \cite{Saffman:2010:RydbergQI}. But
the mechanism of dipole blockade is limited by the radius of dipole-dipole
interaction and can suffer from the decoherence problems arising because of the
strong dependence of dipole-dipole interaction on spatial distance between the
interacting atoms. Besides, another blockade mechanism was recently proposed
based on the dependence of Raman transition frequency on the number of photons
in signal field \cite{Shahriar:2007:atomic-ensemblesQCC} that still remains
rather complicated for experimental realization. We have also proposed a new
decoherence free blockade mechanism based on the Lamb shift in quantum
electrodynamics cavity with additional micro-resonators
\cite{Moiseev:2010:multi-ensembleQC,Andrianov-Moiseev:2011:swapping-gates,Moiseev:2011:multi-ensembleQED,
Andrianov-Moiseev:2012:QC-on-MaC}. Rapid development of micro-resonators
physics and technology
\cite{Duan-Kimble:2004:PhotonicQC,Aoki:2006:strong-coupling,Majer:2007:CoupledQubits}
makes this blockade mechanism very promising though not so simple for
experimental realization.

In this paper, we demonstrate how both single qubit and two qubit gates with
logical encoding of qubits
\cite{Imamoglu:1999:quantum-dot-spins,schuch-2003-67,DiVincenzo:2000:Exchange-Interaction,Bacon:2000:Fault-Tolerant,
Kempe:2001:decoherence-free-computation,Kempe:2001:Encoded-Universality,Kempe:2002:Exact-gate-sequences}
can be realized in natural way on MaC ensembles in the QED cavity without
additional micro-cavities by using only some definite operations of swapping by
excitations between MaC ensembles. Here, we use the encoding of qubits by pairs
of MaC ensembles. In this context the operation of excitation swapping between
such two ensembles corresponds to the NOT gate and swapping operation
controlled by photon from control qubit corresponds to CNOT operation. We
consider physical realization of such quantum gates and show that they can be
used for creation of the quantum computer satisfying the necessary DiVincenzo
criteria \cite{Levy:2002:SpinPairs}.

\section{Excitation swapping on three level systems}

Let's consider a set of three-level MaC ensembles (nodes) posted in the common
electrodynamic cavity (Fig. \ref{Figure1}). Equalizing of resonance transitions
frequencies by the external control electric/magnetic fields leads to fast
excitation swapping between the atoms of nodes by the interaction via the field
of virtual photons that can be described by an effective atomic Hamiltonian
\cite{Moiseev-Andrianov:2012:Photon-echo-quantum-RAM}.
 In order to find this
interaction for the large number of atoms, we start from the initial
Hamiltonian $H=H_{0} +H_{1} $, where $H_{0} =H_{d} +H_{f} $ is the main
Hamiltonian and $H_{1} =H_{d-f} $ is the perturbation Hamiltonian. Here, $H_{d}
=\sum\limits_{m=1}^{2}\sum\limits_{\mu =1}^{3}\sum\limits_{j_{m} =1}^{N_{m}
}\varepsilon _{m}^{\mu } S_{\mu \mu }^{j_{m} } $ is Hamiltonian of atoms in
nodes 1 and 2 in the terms of operator generators $S_{\mu \mu }^{j_{m} } $ of
SU(3) group, where $N_m$ is a number of atoms in $m$-th node, $\varepsilon
_{m}^{\mu } $ is an energy of level $\mu $ in the $m$-th node, and $H_{f}
=\sum\limits_{\alpha }^{1,2,3}\hbar \omega _{k_{\alpha } } a_{k_{\alpha } }^{+}
a_{k_{\alpha } }  $ is Hamiltonian of photons, where $\omega _{k_{\alpha } } $
is the frequency of photons with wave vector $k_{\alpha } $, $a_{k_{\alpha }
}^{+} $ and $a_{k_{\alpha } } $ are operators of creation and annihilation for
photons in modes 1, 2 and 3 (Fig. \ref{Figure1}). For the interaction of
photons with atoms in nodes 1 and 2 $H_{r-a} =H_{r-a}^{\left(1\right)}
+H_{r-a}^{\left(2\right)} =H_{21} +H_{31} +H_{32} $ we have the following
expressions on corresponding atomic transitions:

\begin{figure}
\centerline{\includegraphics[scale=1]{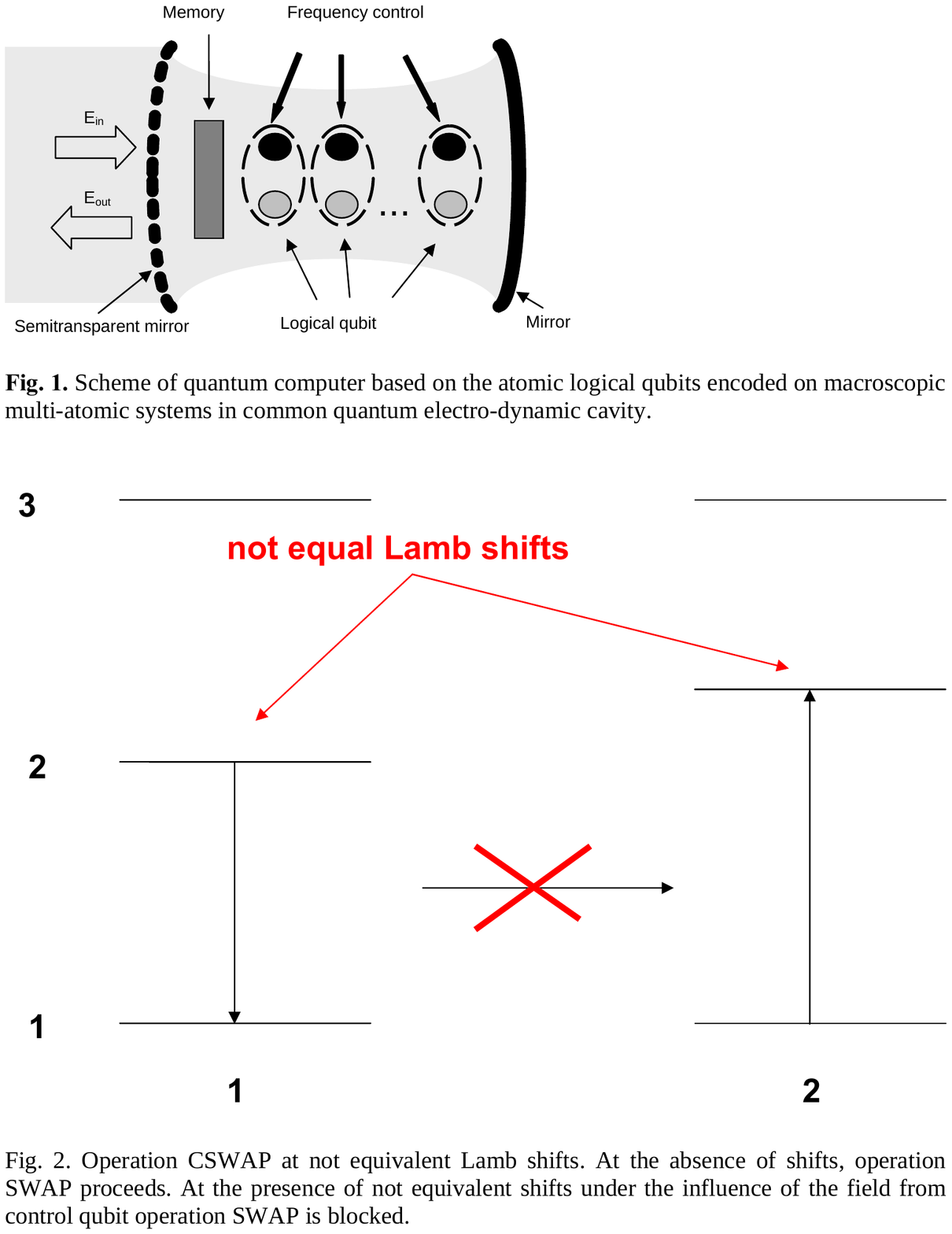}}
\vspace*{8pt}
  \caption{Scheme of quantum computer based on the atomic logical qubits
encoded on macroscopic multi-atomic systems in common quantum electrodynamic
cavity.}
  \label{Figure1}
\end{figure}

\begin{equation}\label{Equation-1_}
H_{21} =\sum\limits_{m}^{1,2}\sum\limits_{j_{m} }\left(g_{21}^{k_{1} } e^{i\vec{k}_{1}
\vec{r}_{j_{m} } } S_{21}^{j_{m} } a_{k_{1} } +g_{21}^{k_{1} *}
e^{-i\vec{k}_{1} \vec{r}_{j_{m} } } S_{12}^{j_{m} } a_{k_{1} }^{+} \right)  ,
\end{equation}

\begin{equation}
H_{32} =\sum\limits_{m}^{1,2}\sum\limits_{j_{m} }\left(g_{32}^{k_{2} } e^{i\vec{k}_{2}
\vec{r}_{j_{m} } } S_{32}^{j_{m} } a_{k_{2} } +g_{32}^{k_{2} *}
e^{-i\vec{k}_{2} \vec{r}_{j_{m} } } S_{23}^{j_{m} } a_{k_{2} }^{+} \right)  ,
\label{Equation-2_}
\end{equation}

\begin{equation}
H_{31} =\sum\limits_{m}^{1,2}\sum\limits_{j_{m} }\left(g_{31}^{k_{3} } e^{i\vec{k}_{3}
\vec{r}_{j_{m} } } S_{31}^{j_{m} } a_{k_{3} } +g_{31}^{k_{3} *}
e^{-i\vec{k}_{3} \vec{r}_{j_{m} } } S_{13}^{j_{m} } a_{k_{3} }^{+} \right)  ,
\label{Equation-3_}
\end{equation}
where $g_{k_{\mu \nu } }^{\left(k_{\alpha } \right)} $ is the interaction
constant for transition from level $\mu$ to level $\nu$, $S_{\mu \nu }^{j_{m} }
$ are the operators of the transitions between the levels $\mu$
$\leftrightarrow$ $\nu $, $\vec{r}_{j_{m} } $ is a radius vector of atoms
$j_{m} $ in nodes $m=1,2$.

In the Appendix \ref{Appendix-A}, we derive an effective Hamiltonian describing
inter-atomic interaction via the virtual photons of QED cavity. The interaction
opens the way for arbitrary quantum manipulation of the atomic ensembles.

In the case when $\Delta _{m}^{\left(1\right)} <<\Delta _{m}^{\left(2\right)}
,\Delta _{m}^{\left(3\right)} $, we can leave in Hamiltonian
\eqref{Equation-10_} only the terms concerning levels 1 and 2 and write it at
$\Delta =\Delta _{1}^{\left(1\right)} =\Delta _{{\rm 2}}^{\left(1\right)} $ as
following:

\begin{equation}\begin{array}{l}H_{s} =\sum\limits_{m}^{1,2}\sum\limits_{j_{m} }\left(\varepsilon
_{m}^{\left(1\right)} S_{11}^{j_{m} } +\varepsilon _{m}^{\left(2\right)}
S_{22}^{j_{m} } \right) +\Omega _{\sigma} \sum\limits_{m=1,2}\sum\limits_{i_{m}
j_{m} }e^{i\vec{k}_{1} \vec{r}_{i_{m} j_{m} } } S_{21}^{i_{m} } S_{12}^{j_{m} }\\
+
\Omega _{\sigma} \sum\limits_{j_{1} j_{2} }\left(e^{i\vec{k}_{1} \vec{r}_{j_{1}
j_{2} } } S_{21}^{j_{1} } S_{12}^{j_{2} } +e^{-i\vec{k}_{1} \vec{r}_{j_{1}
j_{2} } } S_{12}^{j_{1} } S_{21}^{j_{2} } \right) ,
\end{array}
\label{Equation-11_}\end{equation}
where $\Omega _{\sigma} ={\left|g_{21}^{k_{3} } \right|^{2}  / \Delta } $.
Here, the first term describes the atomic energies in the nodes $m = 1$ and 2,
the second term is the energy of atomic excitation swapping inside the nodes,
the third term is the energy of atomic excitation swapping by virtual photons
between various nodes.

 Let's introduce a finite set of collective basic states of all possible Hilbert
 states of the multi-atomic systems in the two nodes:
 ${\left| \psi  \right\rangle} _{1} ={\left| 0 \right\rangle} _{1} {\left| 0
\right\rangle} _{2} $, ${\left| \psi  \right\rangle} _{2} ={\left| 1
\right\rangle} _{1} {\left| 0 \right\rangle} _{2} $, ${\left| \psi
\right\rangle} _{3} ={\left| 0 \right\rangle} _{1} {\left| 1 \right\rangle}
_{2} $, ${\left| \psi  \right\rangle} _{4} ={\left| 1 \right\rangle} _{1}
{\left| 1 \right\rangle} _{2} $, ${\left| \psi  \right\rangle} _{5} ={1
\mathord{\left/{\vphantom{1 \sqrt{2} }}\right.\kern-\nulldelimiterspace}
\sqrt{2} } \left\{{\left| 2 \right\rangle} _{1} {\left| 0 \right\rangle} _{2}
+{\left| 0 \right\rangle} _{1} {\left| 2 \right\rangle} _{2} \right\}$,
${\left| \psi  \right\rangle} _{6} ={1 \mathord{\left/{\vphantom{1 \sqrt{2}
}}\right.\kern-\nulldelimiterspace} \sqrt{2} } \left\{{\left| 2 \right\rangle}
_{1} {\left| 0 \right\rangle} _{2} -{\left| 0 \right\rangle} _{1} {\left| 2
\right\rangle} _{2} \right\}$, where ${\left| 0 \right\rangle} _{1,2} =\prod
_{j_{1,2} }^{N_{1,2} }{\left| g \right\rangle} _{j_{1,2} }  $, ${\left| 1
\right\rangle} _{1,2}$ =\linebreak $(1/\sqrt{N_{1,2} } )\sum\limits_{j_{1,2}
=1}^{N_{1,2} }{\left| e \right\rangle} _{j_{1,2} }  \prod _{l_{1,2} \ne j_{1,2}
}^{N_{1,2} }{\left| g \right\rangle} _{j_{1,2} }  $ and ${\left| 2
\right\rangle} _{1,2} =\sqrt{2/N_{1,2} (N_{1,2} -1)} $ \linebreak
$\cdot\sum\limits_{j_{1,2} >f_{1,2} }^{N_{1,2} }{\left| e \right\rangle}
_{j_{1,2} } {\left| e \right\rangle} _{f_{1,2} } \prod _{l_{1,2} \ne j_{1,2}
,f_{1,2} }^{N_{1,2} }{\left| g \right\rangle} _{j_{1,2} } $. Here for
convenience, the notations: ${\left| g \right\rangle} _{j_{1,2} } $ and
${\left| e \right\rangle} _{j_{1,2} } $ are introduced for the ground and the
first excited states of atom $j_{1,2}$.  It is worth noting that the atomic
states ${\left| \psi  \right\rangle} _{2} $ and ${\left| \psi \right\rangle}
_{3} $ can be prepared by the interaction with single photon states of the
external photon sources. Using these collective states, we find with the
initial state of two qubits of the atoms in two nodes

\begin{equation}{\left| \Psi _{in} \left(0\right) \right\rangle} =\left\{\alpha _{1} {\left| 0
\right\rangle} _{1} +\beta _{1} {\left| 1 \right\rangle} _{1}
\right\}\left\{\alpha _{2} {\left| 0 \right\rangle} _{2} +\beta _{2} {\left| 1
\right\rangle} _{2} \right\},
\label{Equation-12_}
\end{equation}
the following unitary evolution of atomic systems under the interaction part of
Hamiltonian \eqref{Equation-11_}:

\begin{equation}
\label{Equation-13_} \begin{array}{l} {{\left| \Psi \left(t\right) \right\rangle} =\alpha _{1} \alpha _{2} {\left| \psi  \right\rangle} _{1} +} \\ {+
\exp \left(-i\Omega _{\sigma } Nt\right)\left\{\beta _{1} \alpha _{2} \left[\cos \left(\Omega _{\sigma } Nt\right){\left| \psi  \right\rangle} _{2} -
i\sin \left(\Omega _{\sigma } Nt\right){\left| \psi  \right\rangle} _{3} \right]\right. +} \\ {\left. \, \, \, \, \, \, \, \, \, \, \, \, \, \, \, \, \, \, \, \, \, \, \, \, \, \, \, \, \, \, \, +\alpha _{1} \beta _{2}
\left[\cos \left(\Omega _{\sigma } Nt\right){\left| \psi  \right\rangle} _{3} -i\sin \left(\Omega _{\sigma } Nt\right){\left| \psi  \right\rangle} _{2} \right]\right\}+} \\ {+
\exp \left(-i2\Omega _{\sigma } Nt\right)\beta _{1} \beta _{2} \left[\cos \left(2\Omega _{\sigma } Nt\right){\left| \psi  \right\rangle} _{4} -i\sin \left(2\Omega _{\sigma } Nt\right){\left| \psi  \right\rangle} _{5} \right],}
\end{array}
\end{equation}
where we have assumed that the quantity of atoms is sufficiently large $N_1=
N_2 = N >> 1$.

 The solution \eqref{Equation-13_} demonstrates availability of two coherent oscillations with frequency $\Omega _{\sigma } N$ for the first pair of states
 ${\left| \psi  \right\rangle} _{2} \leftrightarrow {\left| \psi  \right\rangle} _{3} $ and with doubled frequency $2\Omega _{\sigma } N$ for the second pair
 ${\left| \psi  \right\rangle} _{4} \leftrightarrow {\left| \psi  \right\rangle} _{5} $.
 Oscillations are strongly accelerated ($N$ and $2N$ times) comparing to the case of quantum oscillations
 of two coupled two-level atoms. 
We have performed a special numeric analysis of the effective Hamiltonian
approach \eqref{Equation-11_} which has demonstrated that large enough spectral
detuning $\Delta>30\sqrt{N}\left|g_{21}^{k_3}\right|$ will provide an error
less than 0.001 for the quantum dynamics presented in Eq. \eqref{Equation-13_}.

\section{Quantum gates}

\subsection{Single-qubit gates}

\begin{figure}
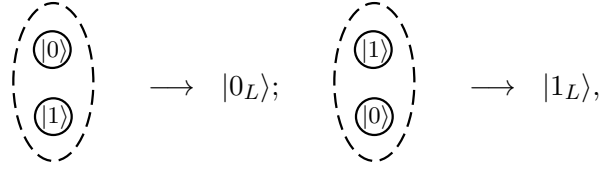

\begin{center}
\begin{tabular}{rclrcl}
\begin{tabular}{r}
\begin{pgfpicture}{21.67mm}{67.71mm}{36.32mm}{92.72mm}
\pgfsetxvec{\pgfpoint{1.00mm}{0mm}}
\pgfsetyvec{\pgfpoint{0mm}{1.00mm}}
\color[rgb]{0,0,0}\pgfsetlinewidth{0.30mm}\pgfsetdash{}{0mm}
\pgfcircle[stroke]{\pgfxy(28.99,75.61)}{2.44mm}
\pgfcircle[stroke]{\pgfxy(28.85,84.53)}{2.44mm}
\pgfsetdash{{2.00mm}{1.00mm}}{0mm}\pgfellipse[stroke]{\pgfxy(28.99,80.22)}{\pgfxy(5.32,0.00)}{\pgfxy(0.00,10.50)}
\pgfputat{\pgfxy(27.24,83.60)}{\pgfbox[bottom,left]{\fontsize{8.54}{10.24}\selectfont $\ket{0}$}}
\pgfputat{\pgfxy(27.32,74.73)}{\pgfbox[bottom,left]{\fontsize{8.54}{10.24}\selectfont $\ket{1}$}}
\end{pgfpicture}
\end{tabular} &
$\longrightarrow$ &
$\ket{0_L}$; &
\begin{tabular}{r}
\begin{pgfpicture}{21.67mm}{67.71mm}{36.32mm}{92.72mm}
\pgfsetxvec{\pgfpoint{1.00mm}{0mm}}
\pgfsetyvec{\pgfpoint{0mm}{1.00mm}}
\color[rgb]{0,0,0}\pgfsetlinewidth{0.30mm}\pgfsetdash{}{0mm}
\pgfcircle[stroke]{\pgfxy(28.99,75.61)}{2.44mm}
\pgfcircle[stroke]{\pgfxy(28.85,84.53)}{2.44mm}
\pgfsetdash{{2.00mm}{1.00mm}}{0mm}\pgfellipse[stroke]{\pgfxy(28.99,80.22)}{\pgfxy(5.32,0.00)}{\pgfxy(0.00,10.50)}
\pgfputat{\pgfxy(27.24,83.60)}{\pgfbox[bottom,left]{\fontsize{8.54}{10.24}\selectfont $\ket{1}$}}
\pgfputat{\pgfxy(27.32,74.73)}{\pgfbox[bottom,left]{\fontsize{8.54}{10.24}\selectfont $\ket{0}$}}
\end{pgfpicture}
\end{tabular} &
$\longrightarrow$ &
$\ket{1_L}$, \\
\end{tabular}
\end{center}
\vspace*{8pt}
  \caption{Pairwise qubit encoding. Small circles denote the processing nodes in the indicated quantum states.}
  \label{Figure:PairwiseEncoding}
\end{figure}

 Here, we introduce an encoding of logical qubits by the pairs of multiatomic systems \cite{Imamoglu:1999:quantum-dot-spins,schuch-2003-67,DiVincenzo:2000:Exchange-Interaction,Bacon:2000:Fault-Tolerant,
Kempe:2001:decoherence-free-computation,Kempe:2001:Encoded-Universality,Kempe:2002:Exact-gate-sequences}
(see Fig. \ref{Figure:PairwiseEncoding}).

 The quantum evolution \eqref{Equation-13_} describes the operation that we
call the \emph{Excitation Transfer}, denoted by $\ET{\theta}$, where $\theta
=\Omega _{\sigma } Nt$ is an arbitrary angle. This operation acts on basis
states $\ket{0}\ket{1}$ and $\ket{1}\ket{0}$ as following:

\begin{equation}
\begin{array}{l}
\ket{0}\ket{1}\longrightarrow \cos{\frac{\theta}{2}}\ket{0}\ket{1} - i\sin{\frac{\theta}{2}}\ket{1}\ket{0}\\
\ket{1}\ket{0}\longrightarrow -i\sin{\frac{\theta}{2}}\ket{0}\ket{1} + \cos{\frac{\theta}{2}}\ket{1}\ket{0}
\end{array}.
\label{Equation-14_}\end{equation}
Thus, for a logical qubit with basis states ${\left| 0 \right\rangle} _{L}
=\ket{0}\ket{1} $ and
 ${\left| 1 \right\rangle} _{L} =\ket{1}\ket{0} $, the $\ET{\theta}$ gate corresponds to the
 rotation $R_{x} \left(\theta \right)$ around the $x$ axis of the Bloch sphere.


So, a quantum computer on multi-atomic systems in the common quantum
electrodynamic cavity (Fig. \ref{Figure1}) can perform a single qubit rotation
in the following way: first, quantum memory is put into the resonance with
designated processing node by equalizing frequencies of this node and the
memory, then the excitation is transferred from the memory to the processing
node by a real photon as described in the paper
\cite{Moiseev-Andrianov:2012:Photon-echo-quantum-RAM}. Next the memory is
decoupled from the node and the node is equalized in frequency with another
processing node of logical qubit so the $\ET{\theta}$ operation takes place
between the two nodes. Finally, the excitation is returned to the memory in the
same way it was transferred from there.

 The proposed architecture also allows us to perform the gate $\PHASE{\chi}$, described as
 following
\begin{equation}
\begin{array}{l}
\ket{0}\ket{1}\longrightarrow e^{-i\chi/2}\ket{0}\ket{1}\\
\ket{1}\ket{0}\longrightarrow e^{i\chi/2}\ket{1}\ket{0}
\end{array},
\end{equation}
where the phases $\chi=\delta\omega\tau$ and 
$\delta\omega$ can be controlled by differing external magnetic (or electric)
fields on the spatially distinct nodes that provides, respectively, different
Zeeman (or Stark) frequency shifts for the two-level atoms localized in the
nodes \cite{Moiseev:2010:multi-ensembleQC}.

When using pairwise encoding of qubits $\PHASE{\chi}$ turns a logical qubit
around the axis $\hat{z}$
, thus corresponding to the logical $R_{\hat{z}}(\chi)$ operator.



That is, $\ET{\theta}$ and $\PHASE{\chi}$ act like orthogonal rotations of the
logical qubits, thus allowing to perform an arbitrary rotation using these
elementary operations.

We note that the described realization of single-qubit gates does't use any
resonant electromagnetic fields that prevents the quantum operations of the
additional excess field noises.

\subsection{Two-qubit gates}

In order to perform universal quantum computation in the Hilbert subspace
corresponding to the pairwise qubit encoding we need an entangling two-qubit
gate such as $\cnot$ gate. In our architecture we will use a
Controlled-ET($\pi$) gate (for short, C(ET)).

\begin{figure}
$$\begin{array}{ccccc}
\begin{array}{l}
\begin{pgfpicture}{2.86mm}{67.71mm}{50.79mm}{96.89mm}
\pgfsetxvec{\pgfpoint{1.00mm}{0mm}}
\pgfsetyvec{\pgfpoint{0mm}{1.00mm}}
\color[rgb]{0,0,0}\pgfsetlinewidth{0.30mm}\pgfsetdash{}{0mm}
\pgfcircle[stroke]{\pgfxy(28.99,75.61)}{2.44mm}
\pgfcircle[stroke]{\pgfxy(28.85,84.53)}{2.44mm}
\pgfsetdash{{2.00mm}{1.00mm}}{0mm}\pgfellipse[stroke]{\pgfxy(28.99,80.22)}{\pgfxy(5.32,0.00)}{\pgfxy(0.00,10.50)}
\pgfputat{\pgfxy(5.79,80.30)}{\pgfbox[bottom,left]{\fontsize{7}{10.24}\selectfont C(ET)}}
\pgfsetdash{}{0mm}\pgfmoveto{\pgfxy(4.86,79.39)}\pgflineto{\pgfxy(17.77,79.39)}\pgflineto{\pgfxy(17.77,82.79)}\pgflineto{\pgfxy(4.86,82.79)}\pgfclosepath\pgfstroke
\pgfmoveto{\pgfxy(17.39,82.84)}\pgfcurveto{\pgfxy(18.15,84.69)}{\pgfxy(19.69,86.11)}{\pgfxy(21.59,86.72)}\pgfcurveto{\pgfxy(22.79,87.11)}{\pgfxy(24.08,87.14)}{\pgfxy(25.30,86.81)}\pgfcurveto{\pgfxy(26.65,86.45)}{\pgfxy(27.85,85.64)}{\pgfxy(28.70,84.53)}\pgfstroke
\pgfmoveto{\pgfxy(28.70,84.53)}\pgflineto{\pgfxy(26.84,86.74)}\pgflineto{\pgfxy(27.00,85.76)}\pgflineto{\pgfxy(26.02,85.60)}\pgflineto{\pgfxy(28.70,84.53)}\pgfclosepath\pgffill
\pgfmoveto{\pgfxy(28.70,84.53)}\pgflineto{\pgfxy(26.84,86.74)}\pgflineto{\pgfxy(27.00,85.76)}\pgflineto{\pgfxy(26.02,85.60)}\pgflineto{\pgfxy(28.70,84.53)}\pgfclosepath\pgfstroke
\pgfmoveto{\pgfxy(17.23,79.34)}\pgfcurveto{\pgfxy(17.22,76.87)}{\pgfxy(18.78,74.65)}{\pgfxy(21.12,73.84)}\pgfcurveto{\pgfxy(22.25,73.45)}{\pgfxy(23.46,73.44)}{\pgfxy(24.64,73.65)}\pgfcurveto{\pgfxy(26.19,73.94)}{\pgfxy(27.65,74.62)}{\pgfxy(28.87,75.63)}\pgfstroke
\pgfmoveto{\pgfxy(28.87,75.63)}\pgflineto{\pgfxy(26.07,74.92)}\pgflineto{\pgfxy(27.02,74.64)}\pgflineto{\pgfxy(26.73,73.69)}\pgflineto{\pgfxy(28.87,75.63)}\pgfclosepath\pgffill
\pgfmoveto{\pgfxy(28.87,75.63)}\pgflineto{\pgfxy(26.07,74.92)}\pgflineto{\pgfxy(27.02,74.64)}\pgflineto{\pgfxy(26.73,73.69)}\pgflineto{\pgfxy(28.87,75.63)}\pgfclosepath\pgfstroke
\pgfcircle[stroke]{\pgfxy(43.47,75.61)}{2.44mm}
\pgfcircle[stroke]{\pgfxy(43.32,84.53)}{2.44mm}
\pgfsetdash{{2.00mm}{1.00mm}}{0mm}\pgfellipse[stroke]{\pgfxy(43.47,80.22)}{\pgfxy(5.32,0.00)}{\pgfxy(0.00,10.50)}
\pgfsetdash{}{0mm}\pgfmoveto{\pgfxy(10.81,83.38)}\pgfcurveto{\pgfxy(13.66,90.89)}{\pgfxy(21.18,95.57)}{\pgfxy(29.18,94.81)}\pgfcurveto{\pgfxy(35.47,94.22)}{\pgfxy(40.94,90.28)}{\pgfxy(43.48,84.50)}\pgfstroke
\pgfmoveto{\pgfxy(10.81,83.38)}\pgflineto{\pgfxy(12.46,85.75)}\pgflineto{\pgfxy(11.56,85.34)}\pgflineto{\pgfxy(11.15,86.24)}\pgflineto{\pgfxy(10.81,83.38)}\pgfclosepath\pgffill
\pgfmoveto{\pgfxy(10.81,83.38)}\pgflineto{\pgfxy(12.46,85.75)}\pgflineto{\pgfxy(11.56,85.34)}\pgflineto{\pgfxy(11.15,86.24)}\pgflineto{\pgfxy(10.81,83.38)}\pgfclosepath\pgfstroke
\pgfputat{\pgfxy(43.42,79.50)}{\pgfbox[bottom,left]{\fontsize{5.69}{6.83}\selectfont \makebox[0pt]{control}}}
\pgfputat{\pgfxy(28.96,79.46)}{\pgfbox[bottom,left]{\fontsize{5.69}{6.83}\selectfont \makebox[0pt]{target}}}
\end{pgfpicture}
\end{array}
& + &
\begin{array}{l}
\begin{pgfpicture}{-1.28mm}{67.71mm}{36.32mm}{92.72mm}
\pgfsetxvec{\pgfpoint{1.00mm}{0mm}}
\pgfsetyvec{\pgfpoint{0mm}{1.00mm}}
\color[rgb]{0,0,0}\pgfsetlinewidth{0.30mm}\pgfsetdash{}{0mm}
\pgfcircle[stroke]{\pgfxy(28.99,75.61)}{2.44mm}
\pgfcircle[stroke]{\pgfxy(28.85,84.53)}{2.44mm}
\pgfsetdash{{2.00mm}{1.00mm}}{0mm}\pgfellipse[stroke]{\pgfxy(28.99,80.22)}{\pgfxy(5.32,0.00)}{\pgfxy(0.00,10.50)}
\pgfputat{\pgfxy(0.96,80.35)}{\pgfbox[bottom,left]{\fontsize{7}{10.24}\selectfont PHASE($\pi$/2)}}
\pgfsetdash{}{0mm}\pgfmoveto{\pgfxy(0.72,79.39)}\pgflineto{\pgfxy(17.77,79.39)}\pgflineto{\pgfxy(17.77,82.79)}\pgflineto{\pgfxy(0.72,82.79)}\pgfclosepath\pgfstroke
\pgfmoveto{\pgfxy(17.39,82.84)}\pgfcurveto{\pgfxy(18.15,84.69)}{\pgfxy(19.69,86.11)}{\pgfxy(21.59,86.72)}\pgfcurveto{\pgfxy(22.79,87.11)}{\pgfxy(24.08,87.14)}{\pgfxy(25.30,86.81)}\pgfcurveto{\pgfxy(26.65,86.45)}{\pgfxy(27.85,85.64)}{\pgfxy(28.70,84.53)}\pgfstroke
\pgfmoveto{\pgfxy(28.70,84.53)}\pgflineto{\pgfxy(26.84,86.74)}\pgflineto{\pgfxy(27.00,85.76)}\pgflineto{\pgfxy(26.02,85.60)}\pgflineto{\pgfxy(28.70,84.53)}\pgfclosepath\pgffill
\pgfmoveto{\pgfxy(28.70,84.53)}\pgflineto{\pgfxy(26.84,86.74)}\pgflineto{\pgfxy(27.00,85.76)}\pgflineto{\pgfxy(26.02,85.60)}\pgflineto{\pgfxy(28.70,84.53)}\pgfclosepath\pgfstroke
\pgfmoveto{\pgfxy(17.23,79.34)}\pgfcurveto{\pgfxy(17.22,76.87)}{\pgfxy(18.78,74.65)}{\pgfxy(21.12,73.84)}\pgfcurveto{\pgfxy(22.25,73.45)}{\pgfxy(23.46,73.44)}{\pgfxy(24.64,73.65)}\pgfcurveto{\pgfxy(26.19,73.94)}{\pgfxy(27.65,74.62)}{\pgfxy(28.87,75.63)}\pgfstroke
\pgfmoveto{\pgfxy(28.87,75.63)}\pgflineto{\pgfxy(26.07,74.92)}\pgflineto{\pgfxy(27.02,74.64)}\pgflineto{\pgfxy(26.73,73.69)}\pgflineto{\pgfxy(28.87,75.63)}\pgfclosepath\pgffill
\pgfmoveto{\pgfxy(28.87,75.63)}\pgflineto{\pgfxy(26.07,74.92)}\pgflineto{\pgfxy(27.02,74.64)}\pgflineto{\pgfxy(26.73,73.69)}\pgflineto{\pgfxy(28.87,75.63)}\pgfclosepath\pgfstroke
\pgfputat{\pgfxy(29.08,79.50)}{\pgfbox[bottom,left]{\fontsize{5.69}{6.83}\selectfont \makebox[0pt]{control}}}
\end{pgfpicture}
\end{array}
& \longrightarrow
&
\begin{array}{l}
\quad \Qcircuit @C=1.0em
@R=1.0em {
 & \ctrl{1} & \qw\\
 & \targ & \qw\\
}
\end{array}
\end{array}
$$
\vspace*{8pt}
  \caption{Logical $\cnot$ gate implemented by the physical $\CET$ gate coupled with $\PHASE{\pi/2}$.}
  \label{Figure:CSWAP-CNOT}
\end{figure}
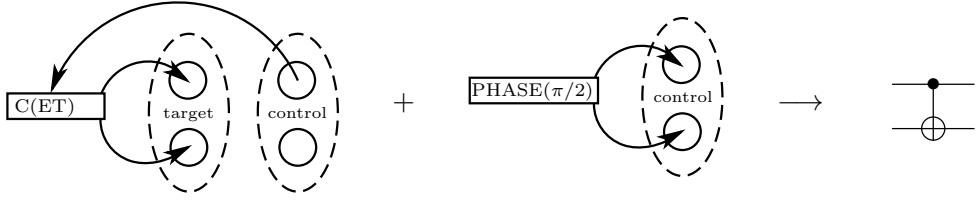

Since the $\ET{\pi}$ operation turns $\ket{01}$ into $\ket{10}$ and backwards
(up to the global phase factor $-i$), thus acting on a pair like the the NOT
gate, $\CET$ implements the logical $\cnot$ gate up to the relative phase
shift, which can be made global by the $\PHASE{\pi/2}$ operation applied to the
control qubit. (see Fig. \ref{Figure:CSWAP-CNOT}).

Formally:
\begin{equation}
\begin{array}{l}
(\PHASE{\pi/2}\otimes I)\cdot\CET\cdot\left(\alpha_1\ket{0_L}\ket{0_L}+\alpha_2\ket{0_L}\ket{1_L}+\alpha_3\ket{1_L}\ket{0_L}+\right.\\
\left.+\alpha_4\ket{1_L}\ket{1_L}\right)=\\
(\PHASE{\pi/2}\otimes I)\cdot\left(\alpha_1\ket{0_L}\ket{0_L}+\alpha_2\ket{0_L}\ket{1_L}-i\alpha_3\ket{1_L}\ket{1_L}-i\alpha_4\ket{1_L}\ket{0_L}\right)=\\
e^{-\frac{\pi}{4}}\left(\alpha_1\ket{0_L}\ket{0_L}+\alpha_2\ket{0_L}\ket{1_L}+\alpha_3\ket{1_L}\ket{1_L}+\alpha_4\ket{1_L}\ket{0_L}\right).
\end{array}
\end{equation}

\begin{figure}
\centerline{\includegraphics[scale=.5]{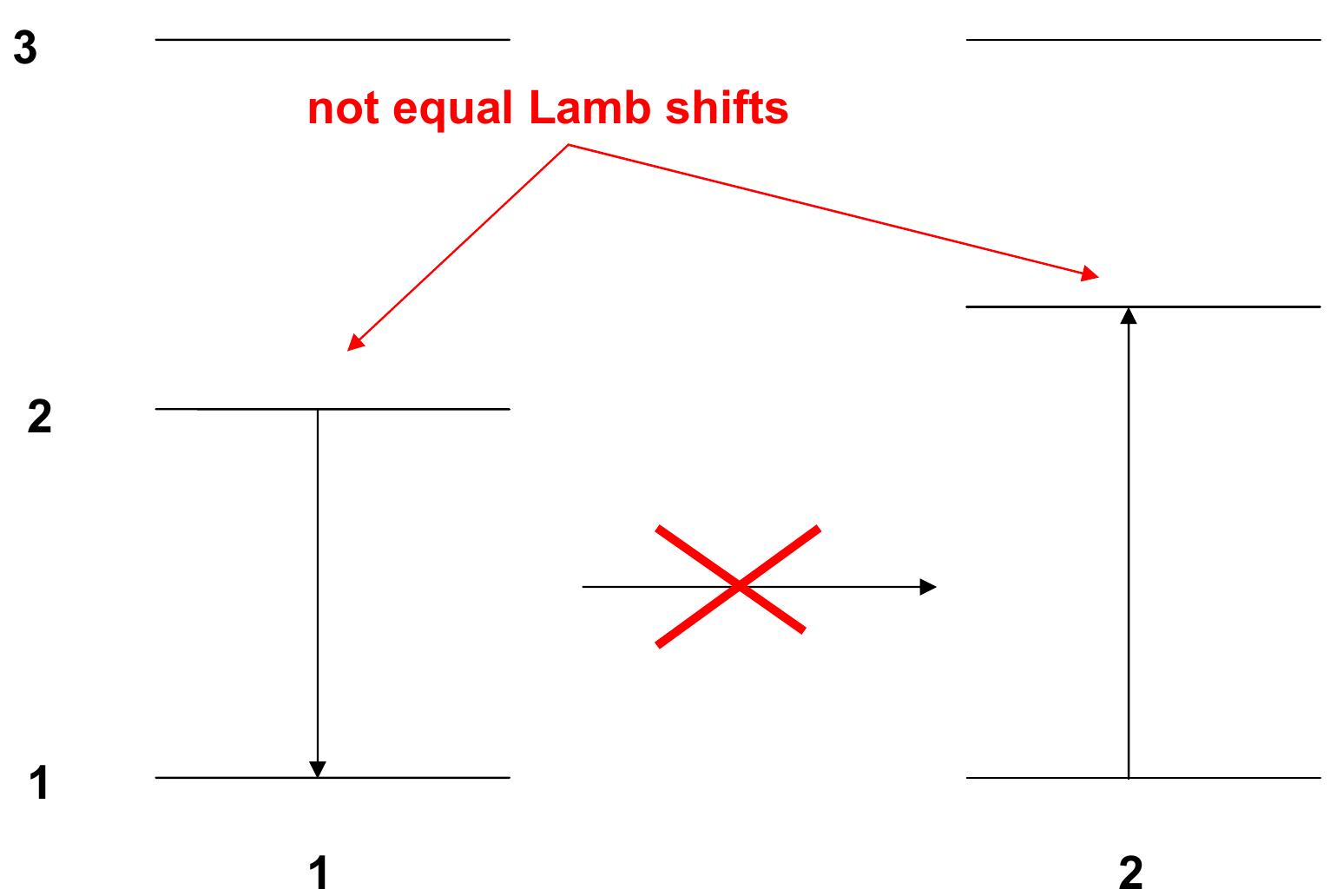}}
\vspace*{8pt}
  \caption{Operation Controlled-ET at not equivalent Lamb shifts. In the absence
of shifts, operation ET proceeds. At the presence of not equivalent shifts
under the influence of the field from control qubit operation ET is blocked.}
  \label{Figure2}
\end{figure}

We propose two schemes for realization of $\CET$ operation on the logical
qubits. The first scheme exploits not equivalent Lamb shift of target logical
qubit under the influence of control qubit field (Fig. \ref{Figure2}). In the
absence of the frequency shifts, ET operation proceeds as it is described in
the previous subsection. The second scheme based on quantum transistor effect
uses a transition of working excitation in target logical qubit at the
additional level under the direct impact of control pair excitation. This
transition takes place if there are atoms of the control qubit
in excited state and the transition then blocks ET operation in the target
logical qubit. Otherwise, if all the atoms of control qubit are in the ground
state, there is no such a transition and ET operation in the target
logical qubit takes place.

Additionally, we will call a compound $\PHASE{\pi/2}$-C(ET) operator a
\emph{Phased Controlled Excitation Transfer} (PCET). Since this operation acts
like Controlled-SWAP for our logical encoding, we will use the C(SWAP) circuit
notation for PCET (as depicted in Figures \ref{Figure:Encoded-Toffoli-gate},
\ref{Figure:Generalized-Controlled-gate-efficient}).

\subsubsection{Gates on nonequivalent Lamb shift}

In order to describe blocking of ET operation at not equivalent Lamb shifts, we
will take into account that if the number of real photons in the cavity
$n_{k_{1} } \neq 0$ then we will have the following Hamiltonian instead of
\eqref{Equation-11_}

\begin{equation}
\label{Equation-15_}
\begin{array}{l} H_{s} =\sum\limits_{m}^{1,2}\sum\limits_{j_{m} \mu }\varepsilon _{m}^{\mu } S_{\mu \mu }^{j_{m} }   +
\sum\limits_{m=1,2}\frac{\left|g_{21}^{\left(m\right)} \left(k_{1} \right)\right|^{2} }{\Delta _{1}^{\left(m\right)} } \sum\limits_{j_{m} }\left(S_{22}^{j_{m} } -S_{11}^{j_{m} } \right)n_{k_{1} } \\
+
\sum\limits_{m=1,2}\frac{\left|g_{21}^{\left(m\right)} \left(k_{1} \right)\right|^{2} }{\Delta _{1}^{\left(m\right)} }
\sum\limits_{i_{m} j_{m} }e^{i\vec{k}_{1} \vec{r}_{i_{m} j_{m} } } S_{21}^{i_{m} } S_{12}^{j_{m} } \\
+\frac{1}{2} \left(\frac{1}{\Delta _{1}^{\left(1\right)} } +
\frac{1}{\Delta _{1}^{\left(2\right)} } \right)\sum\limits_{j_{1} j_{2} }\left(g_{21}^{\left(1\right)} \left(k_{1} \right)g_{21}^{\left(2\right)} \left(k_{1} \right)^{*} e^{i\vec{k}_{1} \vec{r}_{j_{1} j_{2} } }
S_{21}^{j_{1} } S_{12}^{j_{2} }\right.\\
+
\left.g_{21}^{\left(1\right)} \left(k_{1} \right)^{*} g_{21}^{\left(2\right)} \left(k_{1} \right)e^{-i\vec{k}_{1} \vec{r}_{j_{1} j_{2} } } S_{12}^{j_{1} } S_{21}^{j_{2} } \right) \, \, . \end{array}
\end{equation}

Now, the wave function of system can be written in the following form:

\begin{equation}\psi
\left(t\right)=\sum\limits_{n}^{0,1}\sum\limits_{l=1}^{6}c_{l}^{\left(n\right)}
\left(t\right)\psi _{l}  {\left| n \right\rangle} ,
\label{Equation-16_}\end{equation}
where $\psi _{i} $ are given in previous section, and for $n=1$ we get

\begin{equation}\psi \left(t,n=1\right)=\sum\limits_{l=1}^{6}c_{l}^{\left(1\right)}
\left(t\right)\psi _{l}  {\left| 1 \right\rangle} .
\label{Equation-17_}\end{equation}

By writing Schrodinger equation for the wave function \eqref{Equation-17_},
we can see that, in analogy to the evolution described by matrix
\eqref{Equation-14_}, transitions between the states ${\left| 1 \right\rangle}
_{1} {\left| 0 \right\rangle} _{2} $ and ${\left| 0 \right\rangle} _{1} {\left|
1 \right\rangle} _{2} $ proceed separately from transitions between other
atomic states.

Let $\Omega _{1} =\frac{\left|g_{k_{1} }^{\left(1\right)} \right|^{2} }{\hbar
^{2} \Delta _{1} } $, $\Omega _{2} =\frac{\left|g_{k_{1} }^{\left(2\right)}
\right|^{2} }{\hbar ^{2} \Delta _{2} } $, $\Omega _{\sigma} =\frac{g_{k_{1}
}^{\left(1\right)} g_{k_{1} }^{\left(2\right)*} }{2\hbar ^{2} }
\left(\frac{1}{\Delta _{1} } +\frac{1}{\Delta _{2} } \right)$, and
$\widetilde{\omega}_1= \omega_1 +2n_{k_1} \Omega_1$, $\widetilde{\omega}_2 =
\omega_2 +2n_{k_1} \Omega_2$. Then the equations for coefficients $A_{2}
=c_{2}^{\left(1\right)} $ and $A_{3} =c_{3}^{\left(1\right)} $ can be written
in the following form:

\begin{equation}\begin{array}{r}\frac{dc_{2} }{dt} =i\left\{\left(\frac{N_{1} }{2} -1\right)\widetilde{\omega}_1+\frac{N_{2} }{2} \widetilde{\omega}_2-N_{1} \Omega _{1} \right\}c_{2}
-i\sqrt{N_{1} N_{2} } \Omega _{\sigma} c_{3} ,
\label{Equation-18_}
\end{array}
\end{equation}

\begin{equation}\begin{array}{r}\frac{dc_{3} }{dt} =i\left\{\frac{N_{1} }{2} \widetilde{\omega}_1+\left(\frac{N_{2} }{2} -1\right)\widetilde{\omega}_2-N_{2} \Omega _{2} \right\}c_{3}
-i\sqrt{N_{1} N_{2} } \Omega _{\sigma} c_{2},
\label{Equation-19_}
\end{array}
\end{equation}
where $N_{1} $ and $N_{2} $ are the number of atoms in nodes 1 and 2. Solution
of equations \eqref{Equation-18_}, \eqref{Equation-19_} is:

\begin{equation}c_{3} =C_{1} e^{i r_{1} t} +C_{2} e^{i r_{2} t}, \label{Equation-20_}\end{equation}
where $C_{1} $ and $C_{2} $ are constant coefficients,

\begin{equation} \label{Equation-21_} \begin{array}{l} r_{1,2} =\frac{1}{2} \left\{\left(N_{1} -1\right)\widetilde{\omega}_1+
\left(N_{2} -1\right)\widetilde{\omega}_2-N_{1} \Omega _{1} -N_{2} \Omega _{2} \right\}  \\
 {\left. \pm \sqrt{\left( \widetilde{\omega}_2-\widetilde{\omega}_1       +
N_{2} \Omega _{2} -N_{1} \Omega _{1} \right)^{2} +4N_{1} N_{2} \Omega _{\sigma}^{2} } \right\}{\rm \; }.}
\end{array}
\end{equation}

If $c_{2} \left(0\right)=1$ and $c_{3} \left(0\right)=0$ we get

\begin{equation} C_{1} =-C_{2} =-\frac{\sqrt{N_{1} N_{2} } \Omega _{\sigma} }{\sqrt{\left(\widetilde{\omega}_2-\widetilde{\omega}_1+N_{2}
\Omega _{2} -N_{1} \Omega _{1} \right)^{2} +4N_{1} N_{2} \Omega _{\sigma}^{2} } } .
\label{Equation-22_}\end{equation}

When relation $\omega _{2} -\omega _{1} +N_{2} \Omega _{2} -N_{1} \Omega _{1}
=0$ holds we have

\begin{equation} C_{1} =-\frac{\sqrt{N_{1} N_{2} } \Omega _{\sigma} }{2\sqrt{N_{1} N_{2} \Omega
_{s}^{2} +n_{k_{1} } ^{2} \left(\Omega _{2} -\Omega _{1} \right)^{2} } } .
\label{Equation-23_}\end{equation}

If $n_{k_{1} }^{2} \left(\Omega _{2} -\Omega _{1} \right)^{2} >>N_{1} N_{2}
\Omega _{\sigma}^{2} $ and $n_{k_{1} } =1$ then $C_{1} =-C_{2} =c_{3}
\left(t\right)\cong 0$ for any moment of time and ET operation does not
proceed. In the case of $\Omega _{2} >>\Omega _{1} $, we have $\Omega _{2}^{2}
>>N_{1} N_{2} \Omega _{s}^{2} $   and at $\Delta _{1} =\Delta _{2} $ we have
$\Omega _{\sigma}^{2} =\Omega _{1} \Omega _{2} $ that leads to the following
condition $\Omega _{2} >>N_{1} N_{2} \Omega _{1} $. The condition puts the
limit on possible number of atoms in the nodes. If $n_{k_{1} } =0$ then $C_{1}
=-\frac{1}{2} $ and $C_{2} =\frac{1}{2} $. Hence,

\begin{equation}c_{2} =e^{\frac{i}{2\hbar } \left(E_{2} +E_{3} \right)t} \cos \left(\sqrt{N_{1} N_{2} } \Omega _{\sigma} t\right), \label{Equation-24_}\end{equation}

\begin{equation}c_{3} =-ie^{\frac{i}{2\hbar } \left(E_{2} +E_{3} \right)t} \sin
\left(\sqrt{N_{1} N_{2} } \Omega _{\sigma} t\right),
\label{Equation-25_}\end{equation}
where

\begin{equation}E_{2} =\hbar \left\{\left(\frac{N_{1} }{2} -1\right)\omega _{1} +\frac{N_{2}
}{2} \omega _{2} -N_{1} \Omega _{1} \right\},
\label{Equation-26_}\end{equation}

\begin{equation} \label{Equation-27_} E_{3} =\hbar \left\{\frac{N_{1} }{2} \omega _{1} -\left(\frac{N_{2} }{2} -1\right)\omega _{2} -N_{2} \Omega _{2} \right\} \end{equation}
are energies of corresponding states.

Thus, the presence of a photon in the cavity ($n_{k_{1} } =1$) can block ET
operation in the target logical qubit. If this photon arrives from the control
qubit, then we have Controlled-ET operation. Equations (24,25) show that for
small values of $\Omega _{1} $, gate rate is $N$ times higher if $N_{1} =N_{2}
=N$. But above shown constraints put a general upper bound on the gate rate
equal to $\Omega _{2} $ not depending on $N$. So, the best choice in this case
is using quantum dots with large magnitude of intrinsic dipole moment leading
to large $\Omega _{2} $.

\subsubsection{Gates on quantum transistor effect}

\begin{figure}
\centerline{\includegraphics[scale=.5]{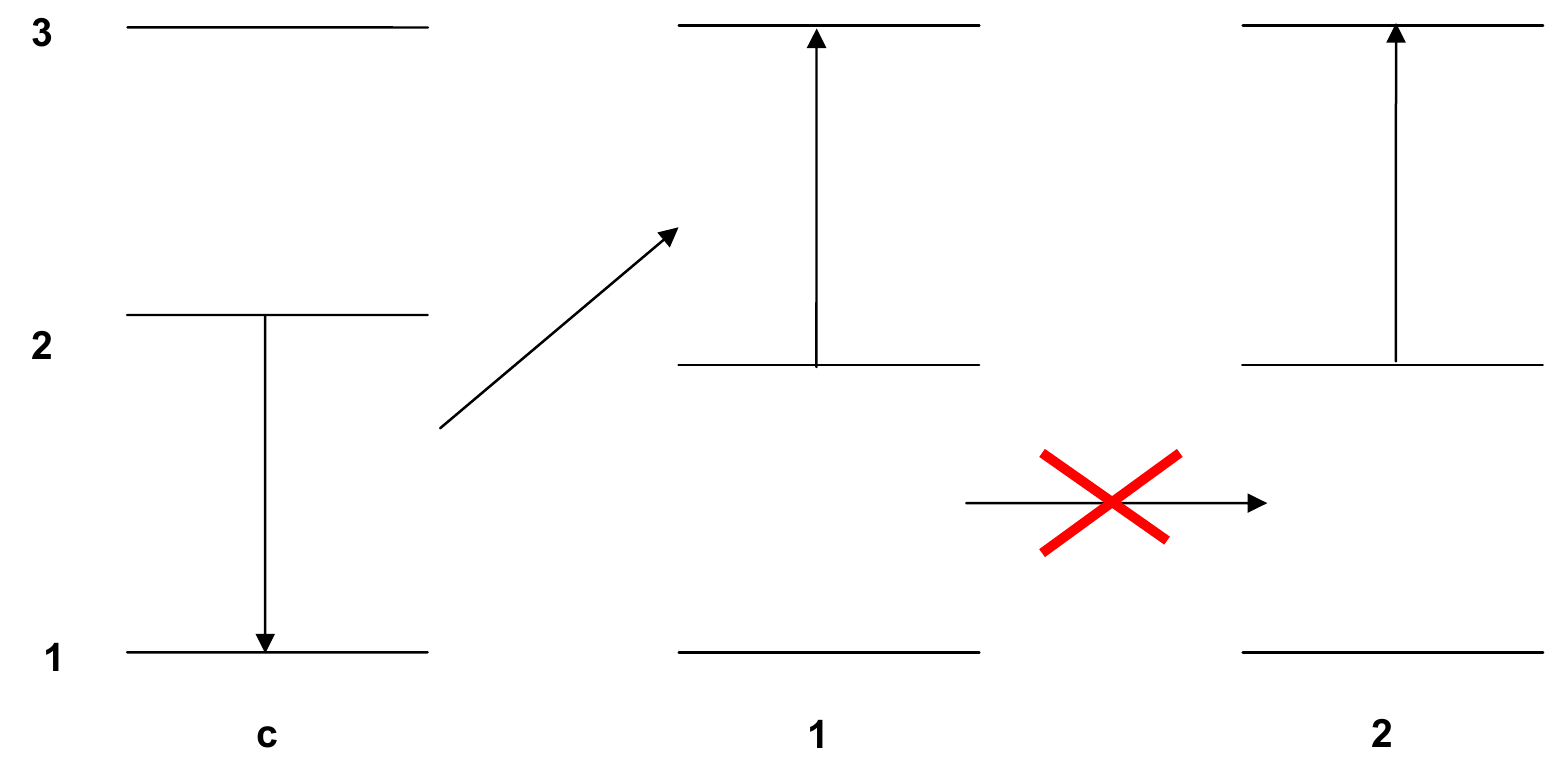}}
\vspace*{8pt}
  \caption{Operation Controlled-ET via quantum transistor effect.
When the atom \textbf{с} of the control logical qubit is not excited operation
ET proceeds between the atoms \textbf{1} and \textbf{2} of the controlled
qubit. When the atom \textbf{с} is excited the ET process at the transition
$2\to 1$ of atom \textbf{с} and $2\to 3$ of atom \textbf{1 }and \textbf{2
}blocks ET operation between atoms \textbf{1} and \textbf{2}.}
  \label{Figure3}
\end{figure}

Another approach to the construction of two-qubit gates on logical qubits is
based on the quantum transistor effect where the ET operation between one of
three-level atomic ensembles of the control qubit at lower quantum transition
and both three-level atomic ensembles of the target qubit at upper transition is
used (Fig. \ref{Figure3}). Here, the frequencies of relevant transitions are
put by the external driving fields in close correspondence to the frequency of
the cavity and the frequencies of the target qubit atomic lower transitions are
detuned far from the resonance. After termination of this process, frequencies
of the target qubit atomic lower transitions are put in close correspondence to
the frequency of the cavity and the frequencies of aforementioned transitions
are detuned far from the resonance. As a result, if the electron of control
qubit atom was in the initial moment of time in the excited state then the ET
process in the target qubits does not proceed since the electrons of its atoms
are on the auxiliary level 3. If, on the contrary, the electron of the control
qubit atom is in the ground state at the initial moment of time then electrons
of the target qubits atoms are partially on the level 2 and ET process takes
place in the target qubits.

Transition frequencies are returned to the previous values after elapsing of
time necessary for ET process in target qubit. As a result, depending on
initial conditions electrons are transferred from the level 3 of the target
qubit atoms to level 2 with the excitation of the control qubit atom in the ET
process or electron rests on level 2 of the target qubits
atom 2 
and operation Controlled-ET is fully executed.

 Hamiltonian describing ET between one of the three-level atomic ensembles
of the control qubit at the lower transition and two of the three-level atomic
ensembles of the target qubits at the upper transition can be written as:

\begin{equation}\begin{array}{r}
H_{ET}^{c\to 1} =\frac{1}{2} \left(\frac{1}{\Delta _{1}^{\left(c\right)} }
+\frac{1}{\Delta _{2}^{\left(1\right)} } \right)\sum\limits_{j_{c} j_{1}
}\left\{g_{32}^{{\left(c\right)} ^{*}} g_{21}^{\left(c\right)}
e^{i\left(\vec{k}_{1} \vec{r}_{j_{c} } -\vec{k}_{2} \vec{r}_{j_{1} } \right)}
S_{21}^{j_{c} } S_{23}^{j_{1} } \right.\\
+
\left.g_{32}^{\left(1\right)}
g_{21}^{{\left(c\right)} ^{*}} e^{-i\left(\vec{k}_{1} \vec{r}_{j_{c} }
-\vec{k}_{2} \vec{r}_{j_{1} } \right)} S_{12}^{j_{c} } S_{32}^{j_{1} } \right\}
,
\label{Equation-28_}
\end{array}
\end{equation}
where
$\frac{1}{\Delta_{0}^{21}}=\frac{1}{\varepsilon_0^{(2)}-\varepsilon_0^{(1)}-\hbar\omega_{k_0}}$,
$\frac{1}{\Delta_{1,2}^{32}}=\frac{1}{\varepsilon_{1,2}^{(3)}-\varepsilon_{1,2}^{(2)}-\hbar\omega_{k_0}}$.
If $\frac{g_{21}^{k_0}}{\Delta_0^{21}}\gg
\frac{g_{32}^{k_0}}{\Delta_{1,2}^{32}}$ we can neglect the ET process at
transition $3\rightarrow2$ during energy transfer from control to target qubit.
We can, also, block the $3\rightarrow2$ ET process executing this energy
transfer to two nodes of the target qubit separately in time detuning
frequencies of target qubit nodes from each other. In the last case, the above
condition is not necessary and we can work with wider set of parameters.

Let's assume for simplicity that atomic ensemble of the control logical qubit
is in the single photonic excited state
$${\left| 1 \right\rangle}
^{\left(A\right)} ={\rm N}^{{\rm -}{{\rm 1} \mathord{\left/{\vphantom{{\rm 1}
2}}\right.\kern-\nulldelimiterspace} 2} } \sum\limits_{j_{c} }e^{i\vec{k}_{1}
\vec{r}_{j_{c} } } {\left| 0 \right\rangle} _{1} ...{\left| 1 \right\rangle}
_{j_{c} }  ...{\left| 0 \right\rangle} _{N}$$ and atomic ensemble of the target
logical qubit transfers as a result of the ET process to two photonic excited
states
$${\left| 2 \right\rangle} ^{\left(m\right)} ={\rm N}^{{\rm -}{{\rm 1}
\mathord{\left/{\vphantom{{\rm 1} 2}}\right.\kern-\nulldelimiterspace} 2} }
\sum _{j_{m} }e^{i\left(\vec{k}_{1} +\vec{k}_{2} \right)\vec{r}_{j_{m} } }
{\left| 0 \right\rangle} _{1_{m} } ...{\left| 2 \right\rangle} _{j_{m} }
...{\left| 0 \right\rangle} _{N_{m} }$$
 of target qubit nodes $m = 1,2$. Then
taking into account the initial state at $t=0$

\begin{equation} \label{Equation-29_}
{\left| \psi (0) \right\rangle} ={\left| 1 \right\rangle} ^{\left(A\right)} \{ \alpha {\left| 0 \right\rangle} ^{\left(1\right)} {\left| 1 \right\rangle} ^{\left(2\right)} +\beta {\left| 1 \right\rangle} ^{\left(1\right)} {\left| 0 \right\rangle} ^{\left(2\right)} \}
, \end{equation}
where $|\alpha |^{2} +|\beta |^{2} =1$, quantum ET-dynamics leads to the
following state of atomic ensembles

\begin{equation}
\label{Equation-30_}
\begin{array}{l} {{\left| \psi (t) \right\rangle} ={\left| 1 \right\rangle} ^{\left(A\right)} \cos \left(\sqrt{N} \Omega t\right)\{ \alpha {\left| 0 \right\rangle} ^{\left(1\right)} {\left| 1 \right\rangle} ^{\left(2\right)} +
\beta {\left| 1 \right\rangle} ^{\left(1\right)} {\left| 0 \right\rangle} ^{\left(2\right)} \} -} \\
{-i\sin \left(\sqrt{N} \Omega t\right){\left| 0 \right\rangle} ^{\left(A\right)} \{ \alpha {\left| 0 \right\rangle} ^{\left(1\right)} {\left|2 \right\rangle} ^{\left(2\right)} +
\beta {\left| 2 \right\rangle} ^{\left(1\right)} {\left| 0 \right\rangle} ^{\left(2\right)} \}
,}
\end{array}
\end{equation}
where

\begin{equation}
\label{Equation-31_}
\Omega =\frac{g_{21}^{k_{0} } g_{32}^{k_{0} } }{2\hbar } \left(\frac{1}{\Delta
_{0}^{12} } +\frac{1}{\Delta _{1}^{23} } \right).
\end{equation}

Here, $\Delta _{1}^{23} =\Delta _{2}^{23} $ and the factor $\sqrt{N}$ indicates
a considerable acceleration of the gate rate comparing with the case of a
single atom.

During QT process (\ref{Equation-30_},\ref{Equation-31_}), also, QT process
between nodes of target qubit at lower transitions partially takes place.
Therefore, it is more convenient practically to achieve energy transfer from
atom of control qubit to atoms of target qubit one by one detuning for a while
resonance between atoms of target qubit in order to avoid excess QT process.

\section{Improved constructions for useful quantum gates}

In this section we demonstrate interesting additional opportunities of the used
atomic systems for accelerated realization of some logical operations. These
opportunities will be possible if we preserve one atomic ensemble in the pure
ground state thus using the states outside our logical Hilbert subspace. Since
due to possible atomic interactions, the atoms could be excited in two possible
states (ground state $\ket{0}$, or $\ket{1}$) this opens a possibility to
efficiently construct a class of useful quantum gates.

\begin{figure}
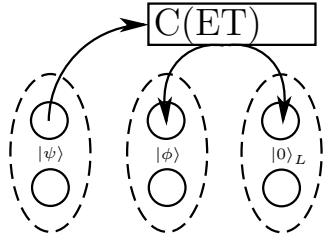

\begin{pgfpicture}{-9.06mm}{67.71mm}{36.32mm}{102.14mm}
\pgfsetxvec{\pgfpoint{1.00mm}{0mm}} \pgfsetyvec{\pgfpoint{0mm}{1.00mm}}
\color[rgb]{0,0,0}\pgfsetlinewidth{0.30mm}\pgfsetdash{}{0mm}
\pgfcircle[stroke]{\pgfxy(28.99,75.61)}{2.44mm}
\pgfcircle[stroke]{\pgfxy(28.85,84.53)}{2.44mm}
\pgfsetdash{{2.00mm}{1.00mm}}{0mm}\pgfellipse[stroke]{\pgfxy(28.99,80.22)}{\pgfxy(5.32,0.00)}{\pgfxy(0.00,10.50)}
\pgfputat{\pgfxy(12.01,95.92)}{\pgfbox[bottom,left]{\fontsize{14.23}{17.07}\selectfont
C(ET)}}
\pgfsetdash{}{0mm}\pgfmoveto{\pgfxy(11.29,94.65)}\pgflineto{\pgfxy(33.50,94.65)}\pgflineto{\pgfxy(33.50,100.14)}\pgflineto{\pgfxy(11.29,100.14)}\pgfclosepath\pgfstroke
\pgfcircle[stroke]{\pgfxy(-1.74,75.66)}{2.44mm}
\pgfcircle[stroke]{\pgfxy(-1.88,84.58)}{2.44mm}
\pgfsetdash{{2.00mm}{1.00mm}}{0mm}\pgfellipse[stroke]{\pgfxy(-1.74,80.27)}{\pgfxy(5.32,0.00)}{\pgfxy(0.00,10.50)}
\pgfputat{\pgfxy(-3.5,79.55)}{\pgfbox[bottom,left]{\fontsize{5.69}{6.83}\selectfont
$\ket{\psi}$}}
\pgfputat{\pgfxy(27.5,79.46)}{\pgfbox[bottom,left]{\fontsize{5.69}{6.83}\selectfont
$\ket{0}_L$}}
\pgfsetdash{}{0mm}\pgfcircle[stroke]{\pgfxy(13.78,75.66)}{2.44mm}
\pgfcircle[stroke]{\pgfxy(13.64,84.58)}{2.44mm}
\pgfsetdash{{2.00mm}{1.00mm}}{0mm}\pgfellipse[stroke]{\pgfxy(13.78,80.27)}{\pgfxy(5.32,0.00)}{\pgfxy(0.00,10.50)}
\pgfputat{\pgfxy(12.2,79.51)}{\pgfbox[bottom,left]{\fontsize{5.69}{6.83}\selectfont
$\ket{\phi}$}}
\pgfsetdash{}{0mm}\pgfmoveto{\pgfxy(-1.96,84.53)}\pgfcurveto{\pgfxy(-1.91,88.35)}{\pgfxy(-0.18,91.96)}{\pgfxy(2.76,94.39)}\pgfcurveto{\pgfxy(5.01,96.25)}{\pgfxy(7.82,97.28)}{\pgfxy(10.73,97.33)}\pgfstroke
\pgfmoveto{\pgfxy(10.73,97.33)}\pgflineto{\pgfxy(7.92,97.99)}\pgflineto{\pgfxy(8.63,97.30)}\pgflineto{\pgfxy(7.94,96.59)}\pgflineto{\pgfxy(10.73,97.33)}\pgfclosepath\pgffill
\pgfmoveto{\pgfxy(10.73,97.33)}\pgflineto{\pgfxy(7.92,97.99)}\pgflineto{\pgfxy(8.63,97.30)}\pgflineto{\pgfxy(7.94,96.59)}\pgflineto{\pgfxy(10.73,97.33)}\pgfclosepath\pgfstroke
\pgfmoveto{\pgfxy(21.96,94.60)}\pgfcurveto{\pgfxy(24.82,95.02)}{\pgfxy(27.61,93.47)}{\pgfxy(28.77,90.83)}\pgfcurveto{\pgfxy(29.29,89.64)}{\pgfxy(29.42,88.34)}{\pgfxy(29.40,87.05)}\pgfcurveto{\pgfxy(29.39,86.24)}{\pgfxy(29.32,85.44)}{\pgfxy(29.19,84.64)}\pgfstroke
\pgfmoveto{\pgfxy(29.19,84.64)}\pgflineto{\pgfxy(30.03,87.40)}\pgflineto{\pgfxy(29.30,86.74)}\pgflineto{\pgfxy(28.63,87.47)}\pgflineto{\pgfxy(29.19,84.64)}\pgfclosepath\pgffill
\pgfmoveto{\pgfxy(29.19,84.64)}\pgflineto{\pgfxy(30.03,87.40)}\pgflineto{\pgfxy(29.30,86.74)}\pgflineto{\pgfxy(28.63,87.47)}\pgflineto{\pgfxy(29.19,84.64)}\pgfclosepath\pgfstroke
\pgfmoveto{\pgfxy(21.27,94.55)}\pgfcurveto{\pgfxy(18.34,95.07)}{\pgfxy(15.43,93.56)}{\pgfxy(14.14,90.88)}\pgfcurveto{\pgfxy(13.58,89.71)}{\pgfxy(13.40,88.40)}{\pgfxy(13.41,87.10)}\pgfcurveto{\pgfxy(13.41,86.22)}{\pgfxy(13.50,85.34)}{\pgfxy(13.67,84.48)}\pgfstroke
\pgfmoveto{\pgfxy(13.67,84.48)}\pgflineto{\pgfxy(14.20,87.32)}\pgflineto{\pgfxy(13.54,86.58)}\pgflineto{\pgfxy(12.80,87.23)}\pgflineto{\pgfxy(13.67,84.48)}\pgfclosepath\pgffill
\pgfmoveto{\pgfxy(13.67,84.48)}\pgflineto{\pgfxy(14.20,87.32)}\pgflineto{\pgfxy(13.54,86.58)}\pgflineto{\pgfxy(12.80,87.23)}\pgflineto{\pgfxy(13.67,84.48)}\pgfclosepath\pgfstroke
\end{pgfpicture}
\vspace*{8pt}
  \caption{Implementing logical AND operation using C(ET).}
  \label{Figure:AND-gate-from-CSWAP}
\end{figure}

The key idea is to use the PCET operation to compute the logical AND operation
in a manner depicted in Fig. \ref{Figure:AND-gate-from-CSWAP}. It acts as
following:

$$\ket{\phi_1\phi_2}\ket{\psi_1\psi_2}\ket{0}\ket{1}\longrightarrow\ket{\phi_1\phi_2}\ket{\widetilde{\psi_1}\psi_2}\ket{\phi_1\mbox{ AND }\psi_1}\ket{1}.$$

It can be easily seen that the first node of the third pair will turn into the
state $\ket{1}$ iff both $\ket{\phi}$ and $\ket{\psi}$ were in the state
$\ket{1}$ (which shows that this operation actually performs the AND of $\phi$
and $\psi$):
$$
(\alpha_1\ket{01}\ket{01}+\alpha_2\ket{01}\ket{10}+\alpha_3\ket{10}\ket{01}+\alpha_4\ket{10}\ket{10})\otimes\ket{01} \longrightarrow
$$
$$
\left(\alpha_1\ket{01}\ket{01}+\alpha_2\ket{01}\ket{10} + \alpha_3\ket{10}\ket{01}\right)\otimes\ket{01}+\alpha_4\ket{10}\ket{00}\ket{11}.
$$

However, this operation extends the logical encoding which we have adopted for
our model and we have to make sure to return the state into our Hilbert
subspace after performing such ``low-level'' operations. In that case we don't
actually need the third qubit as a pair of nodes, all we need is a single node
in the ground state $\ket{0}$.

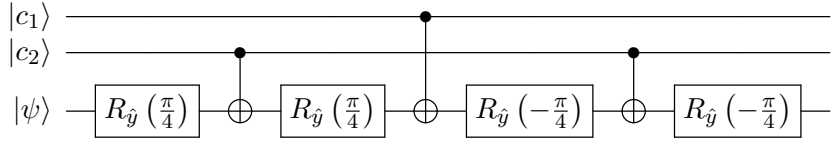
\begin{figure}
$$
\begin{array}{l}\quad\quad~ \Qcircuit
@C=1.0em @R=1.0em {
\lstick{\ket{c_1}} & \qw & \qw & \qw & \ctrl{2} & \qw & \qw & \qw & \qw\\
\lstick{\ket{c_2}} & \qw & \ctrl{1} & \qw & \qw & \qw & \ctrl{1} & \qw & \qw\\
\lstick{\ket{\psi}} & \gate{R_{\hat{y}}\left(\frac{\pi}{4}\right)} & \targ & \gate{R_{\hat{y}}\left(\frac{\pi}{4}\right)} & \targ & \gate{R_{\hat{y}}\left(-\frac{\pi}{4}\right)} & \targ & \gate{R_{\hat{y}}\left(-\frac{\pi}{4}\right)} & \qw\\
}\end{array}
$$
\vspace*{8pt}
  \caption{Circuit implementing Toffoli gate up to relative phase factor.}
  \label{Figure:Toffoli-gate-standard}
\end{figure}

As an example let's consider one of the most commonly used quantum gate -- the
Toffoli gate (also known as CCNOT). Using the universal set of $\cnot$ and
single qubit gates it can be implemented up to relative phase factor using the
circuit from \cite{Nielsen-Chuang:2000:QC} (see Fig.
\ref{Figure:Toffoli-gate-standard}).

\begin{figure}
$$
\begin{array}{l}\quad\quad \Qcircuit  @C=1.0em @R=1.0em {
\lstick{\begin{array}{r} \\\left.\ket{c_{1,L}}\right\{\end{array}} & \ctrl{2} & \qw\\
 & \qw & \qw\\
\lstick{\begin{array}{r} \\\left.\ket{c_{2,L}}\right\{\end{array}} & \ctrl{2} & \qw\\
 & \qw & \qw\\
\lstick{\begin{array}{r} \\\left.\ket{\psi_L}\right\{\end{array}} & \qswap \qwx[1]& \qw\\
 & \qswap & \qw\\
\lstick{\ket{0}} & \qw & \qw\\
}\end{array} =
\begin{array}{l}\quad\quad\quad\quad \Qcircuit
@C=1.0em @R=1.0em {
\lstick{\begin{array}{r} \\\left.\ket{c_{1,L}}\right\{\end{array}} & \ctrl{2} & \qw & \ctrl{2} & \qw\\
 & \qw & \qw & \qw & \qw\\
\lstick{\begin{array}{r} \\\left.\ket{c_{2,L}}\right\{\end{array}} & \qswap \qwx[4] & \qw & \qswap \qwx[4] & \qw\\
 & \qw & \qw & \qw & \qw\\
\lstick{\begin{array}{r} \\\left.\ket{\psi_L}\right\{\end{array}} & \qw & \qswap \qwx[1] & \qw & \qw\\
 & \qw & \qswap & \qw & \qw\\
\lstick{\ket{0}} & \qswap & \ctrl{-1} & \qswap & \qw\\
}\end{array}
\longrightarrow
\begin{array}{l}\quad\quad\quad \Qcircuit  @C=1.0em @R=1.0em {
\lstick{\ket{c_{1,L}}} & \ctrl{1} & \qw\\
\lstick{\ket{c_{2,L}}} & \ctrl{1} & \qw\\
\lstick{\ket{\psi_L}} & \targ & \qw\\
}\end{array}
$$
\vspace*{8pt}
  \caption{Efficient implementation of the encoded Toffoli gate using the
physical PCET operations (depicted using Controlled-SWAP notation).}
  \label{Figure:Encoded-Toffoli-gate}
\end{figure}
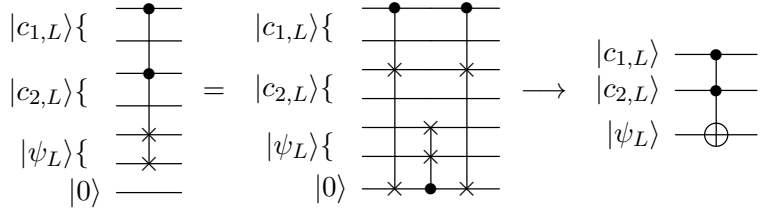

We can construct a more efficient \quotes{low-level} circuit 
using PCET gates and an extra processing node in the state $\ket{0}$. The
circuit in Fig. \ref{Figure:Encoded-Toffoli-gate} checks if both of control
qubits (logical) are in the state $\ket{1}$ and stores the result in an
auxiliary qubit (physical, single node). Afterwards, it performs the PCET of
the third pair controlled by the auxiliary qubit, so the third qubit
$\ket{\psi_L}$ is flipped iff both of $\ket{c_{1,L}}$ and $\ket{c_{2,L}}$ were
in the state $\ket{1_L}$, which is exactly the Toffoli gate. Finally, our
circuit uncomputes the first operation, returning the auxiliary qubit into
state $\ket{0}$ and the whole system into the Hilbert subspace corresponding to
our logical qubit encoding.


This approach can be generalized to improve constructions of the general
controlled gate $C^t(U)$, defined by the following equation in
\cite{Nielsen-Chuang:2000:QC}:
\begin{equation}
C^t(U)\ket{c_1c_2\ldots c_t}\ket{\psi}=\ket{c_1c_2\ldots c_t}U^{c_1\cdot c_2\cdots c_t}\ket{\psi}.
\end{equation}

\begin{figure}
$$
\begin{array}{l}\quad\quad \Qcircuit  @C=1.0em @R=1.0em {
\lstick{\ket{c_1}} & \ctrl{1} & \qw\\
\lstick{\ket{c_2}} & \ctrl{1} & \qw\\
\lstick{\ket{c_3}} & \ctrl{1} & \qw\\
\lstick{\ket{c_4}} & \ctrl{1} & \qw\\
\lstick{\ket{\psi}} & \gate{U} & \qw\\
}\end{array} = \begin{array}{l}\quad\quad
\Qcircuit  @C=1.0em @R=1.0em {
\lstick{\ket{c_1}} & \ctrl{1} & \qw & \qw & \qw & \qw & \qw & \ctrl{1} & \qw\\
\lstick{\ket{c_2}} & \ctrl{4} & \qw & \qw & \qw & \qw & \qw & \ctrl{4} & \qw\\
\lstick{\ket{c_3}} & \qw & \ctrl{1} & \qw & \qw & \qw & \ctrl{1} & \qw & \qw\\
\lstick{\ket{c_4}} & \qw & \ctrl{3}& \qw & \qw & \qw & \ctrl{3} & \qw & \qw\\
\lstick{\ket{\psi}} & \qw & \qw & \qw & \gate{U} & \qw & \qw & \qw & \qw\\
\lstick{\ket{0}} & \targ & \qw & \ctrl{1} & \qw & \ctrl{1} & \qw & \targ & \qw\\
\lstick{\ket{0}} & \qw & \targ & \ctrl{1} & \qw & \ctrl{1} & \targ & \qw & \qw\\
\lstick{\ket{0}} & \qw & \qw & \targ & \ctrl{-3} & \targ & \qw & \qw & \qw\\
}
\end{array}
$$
\vspace*{8pt}
  \caption{Implementation of the gate $U$ controlled by 4 qubits.}
  \label{Figure:Generalized-Controlled-gate-standard}
\end{figure}
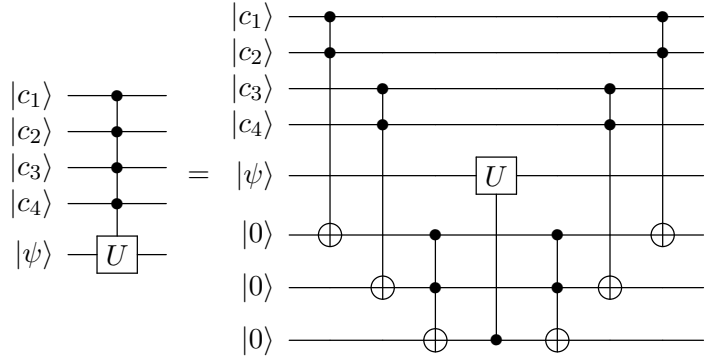

The usual construction exploiting ancillary qubits in state $\ket{0}$ is the
one demonstrated for $t=4$ in Fig.
\ref{Figure:Generalized-Controlled-gate-standard}.

Such a circuit for $C^t(U)$ requires $2(t-1)$ Toffoli gates and one
controlled-$U$ operation plus $t-1$ qubits (initially in the state $\ket{0}$)
for temporary storage. As we already know each encoded Toffoli gate can be
implemented using three PCET gates and additional node (\quotes{half} of the
logical qubit) in the state $\ket{0}$. Of course, for non-parallel Toffoli
gates we can use the same ancillary node, for it remains in the state
$\ket{0}$.

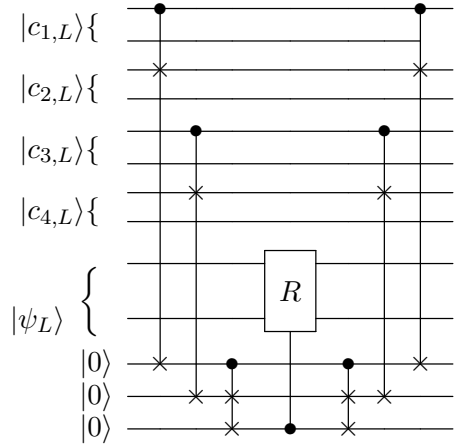
\begin{figure}
\[\quad\quad\quad
\Qcircuit  @C=1.0em @R=1.0em {
\lstick{\begin{array}{r} \\\left.\ket{c_{1,L}}\right\{\end{array}} & \ctrl{2} & \qw & \qw & \qw & \qw & \qw & \ctrl{2} & \qw\\
 & \qw & \qw & \qw & \qw & \qw & \qw & \qw & \\
\lstick{\begin{array}{r} \\\left.\ket{c_{2,L}}\right\{\end{array}} & \qswap\qwx[8] & \qw & \qw & \qw & \qw & \qw & \qswap\qwx[8] & \qw\\
 & \qw & \qw & \qw & \qw & \qw & \qw & \qw & \qw\\
\lstick{\begin{array}{r} \\\left.\ket{c_{3,L}}\right\{\end{array}} & \qw & \ctrl{2} & \qw & \qw & \qw & \ctrl{2} & \qw & \qw\\
 & \qw & \qw & \qw & \qw & \qw & \qw & \qw & \qw\\
\lstick{\begin{array}{r} \\\left.\ket{c_{4,L}}\right\{\end{array}} & \qw & \qswap\qwx[5]& \qw & \qw & \qw & \qswap\qwx[5] & \qw & \qw\\
 & \qw & \qw & \qw & \qw & \qw & \qw & \qw & \qw\\
\lstick{\begin{array}{r}\\\\\left.\begin{array}{r}\\ \ket{\psi_L}\end{array}\right\{\end{array}} & \qw & \qw & \qw & \multigate{1}{R} & \qw & \qw & \qw & \qw\\
 & \qw & \qw & \qw & \ghost{R} & \qw & \qw & \qw & \qw\\
\lstick{\ket{0}} & \qswap & \qw & \ctrl{1} & \qw & \ctrl{1} & \qw & \qswap & \qw\\
\lstick{\ket{0}} & \qw & \qswap & \qswap\qwx[1] & \qw & \qswap\qwx[1] & \qswap & \qw & \qw\\
\lstick{\ket{0}} & \qw & \qw & \qswap & \ctrl{-3} & \qswap & \qw & \qw & \qw\\
}
\]
\vspace*{8pt}
  \caption{Improved implementation of the gate $U$ controlled by 4 qubits.}
  \label{Figure:Generalized-Controlled-gate-efficient}
\end{figure}

On the other hand, we can use the scheme of Fig.
\ref{Figure:Generalized-Controlled-gate-efficient} for implementing the logical
$C^t(U)$ gate. Here, we have made use of $2(t-1)$ PCETs (which are implemented
by two elementary gates in our model) and $t-1$ ancillary processing nodes
instead of full logical qubits.

\section{Conclusion}

Thus, we have introduced a logical encoding of qubits for multi-atomic systems
and demonstrated its advantages for construction of universal quantum
computations. Here, the logical qubits are realized by the pairs of atomic
ensembles that provides a high rate of single qubit gates without  excess
noises on the resonant atomic frequencies. We have elaborated this approach
for the construction both of single and two-qubit gates based on transfer of
excitation between the atomic ensembles via exchange of virtual photons in the
QED cavity.  This opens rich possibilities
for quantum processing over the large number of qubits. The proposed
architecture is scalable due to the free of decoherence nature of interaction
between the multi-atomic ensembles in the QED cavity. We have also shown a
possibility of using additional quantum states (outside of the Hilbert subspace
generated by the logical encoding) of atomic ensembles for multiple
acceleration of some complex controlled operations,
 which are at the core of many
quantum algorithms.

We have considered two types of two qubit gates: the one is based on
nonequivalent Lamb shifts of collective transitions in the multi-atomic systems
and the other one
- on quantum transistor effect. The quantum transistor effect uses control of resonant transfer
of excitation between target atomic ensembles due to the reversible change of atomic levels population.  
 Quantum transistor effect
is very promising for the construction of quantum computer especially taking
into account comparatively long decoherence times of natural atoms. Both
approaches can be realized by using existing  facilities in quantum optics and
in microwave QED technology.

\bibliography{references,physics}

\newpage

\begin{appendix}

\section{}\label{Appendix-A}

Let's perform a unitary transformation of the Hamiltonian $H_{s} =e^{-s}
He^{s}$ that yields the following result in a second order of the perturbation
theory:

\begin{equation}
H_{s} =H_{0} +\frac{1}{2} \left[H_{1} ,s\right], \label{Equation-3a}
\end{equation} when the
relation

\begin{equation}\label{Equation-4_}
H_{1} +\left[H_{0} ,s\right]=0,
\end{equation}
holds.

Using the relation \eqref{Equation-4_}, we obtain

\begin{equation} \label{Equation-5_}
\begin{array}{l} {s=}  \sum\limits_{j_{1} }\left(\alpha _{1}^{\left(1\right)} g_{21}^{k_{1} } e^{i\vec{k}_{1} \vec{r}_{j_{1} } } S_{21}^{j_{1} } a_{k_{1} } +\beta _{1}^{\left(1\right)}
g_{21}^{k_{1} *} e^{-i\vec{k}_{1} \vec{r}_{j_{1} } } S_{12}^{j_{1} } a_{k_{1} }^{+} \right)\\
+
\sum\limits_{j_{2} }\left(\alpha _{2}^{\left(1\right)} g_{21}^{k_{1} } e^{i\vec{k}_{1} \vec{r}_{j_{2} } } S_{21}^{j_{2} } a_{k_{1} } +
\beta _{2}^{\left(1\right)} g_{21}^{k_{1} *} e^{-i\vec{k}_{1} \vec{r}_{j_{2} } } S_{12}^{j_{2} } a_{k_{1} }^{+} \right) \\
+
\sum\limits_{j_{1} }\left(\alpha _{1}^{\left(2\right)} g_{32}^{k_{2} } e^{i\vec{k}_{2} \vec{r}_{j_{1} } }
S_{32}^{j_{1} } a_{k_{2} } +\beta _{1}^{\left(2\right)} g_{32}^{k_{2} *} e^{-i\vec{k}_{2} \vec{r}_{j_{1} } } S_{23}^{j_{1} } a_{k_{2} }^{+} \right)\\
+
\sum\limits_{j_{2} }\left(\alpha _{2}^{\left(2\right)} g_{32}^{k_{2} } e^{i\vec{k}_{2}
\vec{r}_{j_{2} } } S_{32}^{j_{2} } a_{k_{2} } +\beta _{2}^{\left(2\right)} g_{32}^{k_{2} *} e^{-i\vec{k}_{2} \vec{r}_{j_{2} } } S_{23}^{j_{2} } a_{k_{2} }^{+} \right)\\
+
\sum\limits_{j_{1} }\left(\alpha _{1}^{\left(3\right)}
g_{31}^{k_{3} } e^{i\vec{k}_{3} \vec{r}_{j_{1} } } S_{31}^{j_{1} } a_{k_{3} } +\beta _{1}^{\left(3\right)} g_{31}^{k_{3} *} e^{-i\vec{k}_{3} \vec{r}_{j_{1} } } S_{13}^{j_{1} } a_{k_{3} }^{+} \right)\\
+
\sum\limits_{j_{2} }\left(\alpha _{1}^{\left(3\right)} g_{31}^{k_{3} } e^{i\vec{k}_{3} \vec{r}_{j_{2} } } S_{31}^{j_{2} } a_{k_{3} } +\beta _{1}^{\left(3\right)} g_{31}^{k_{3} *} e^{-i\vec{k}_{3} \vec{r}_{j_{2} } } S_{13}^{j_{2} }
a_{k_{3} }^{+} \right) \, \, , \end{array}
\end{equation}

where

\begin{equation}\alpha _{1,2}^{\left(1\right)} =-\beta _{1,2}^{\left(1\right)}
=-\frac{1}{\varepsilon _{1,2}^{\left(2\right)} -\varepsilon
_{1,2}^{\left(1\right)} -\hbar \omega _{k_{3} } } ,
\label{Equation-6_}\end{equation}

\begin{equation}\alpha _{1,2}^{\left(2\right)} =-\beta _{1,2}^{\left(2\right)}
=-\frac{1}{\varepsilon _{1,2}^{\left(3\right)} -\varepsilon
_{1,2}^{\left(2\right)} -\hbar \omega _{k_{3} } } ,
\label{Equation-7_}\end{equation}

\begin{equation}\alpha _{1,2}^{\left(3\right)} =-\beta _{1,2}^{\left(3\right)}
=-\frac{1}{\varepsilon _{1,2}^{\left(3\right)} -\varepsilon
_{1,2}^{\left(1\right)} -\hbar \omega _{k_{3} } } .
\label{Equation-8_}\end{equation}

 Substituting expressions \eqref{Equation-4_} and \eqref{Equation-5_} into \eqref{Equation-3a}, we get

\begin{equation}
\label{Equation-9_} \begin{array}{l}
{H_{s} =} \sum\limits_{m=1,2}\frac{\left|g_{21}^{k_{3} } \right|^{2} }{\Delta _{m}^{\left(1\right)} } \sum\limits_{i_{m} j_{m} }e^{i\vec{k}_{1} \vec{r}_{i_{m} j_{m} } } S_{21}^{i_{m} } S_{12}^{j_{m} } +
\sum\limits_{m=1,2}\frac{\left|g_{21}^{k_{3} } \right|^{2} }{\Delta _{m}^{\left(1\right)} }\sum\limits_{j_{m} }\left(S_{22}^{j_{m} } -S_{11}^{j_{m} } \right)n_{k_{1} } \\
+
 \frac{1}{2} \left(\frac{1}{\Delta _{1}^{\left(1\right)} } +\frac{1}{\Delta _{2}^{\left(1\right)} } \right)\left|g_{21}^{k_{1} } \right|^{2}
\sum\limits_{j_{1} j_{2} }\left(e^{i\vec{k}_{1} \vec{r}_{j_{1} j_{2} } } S_{21}^{j_{1} } S_{12}^{j_{2} } +h.c.\right) \\
+
\frac{1}{2} \sum\limits_{m=1,2}\frac{1}{\Delta _{m}^{\left(1\right)} }
\sum\limits_{j_{1} }\left(g_{32}^{k_{2} } g_{21}^{k_{1} } e^{i\left(\vec{k}_{1} +\vec{k}_{2} \right)\vec{r}_{j_{1} } } S_{31}^{i_{1} } a_{k_{2} } a_{k_{1} } +h.c.\right) \\
+
\frac{1}{2} \sum\limits_{m=1,2}\frac{1}{\Delta _{m}^{\left(1\right)} }
\sum\limits_{j_{2} }\left(g_{31}^{k_{3} } g_{21}^{k_{1} *} e^{i\left(\vec{k}_{3} -\vec{k}_{1} \right)\vec{r}_{j_{2} } } S_{32}^{i_{2} } a_{k_{3} } a_{k_{1} }^{+} +h.c.\right)\\
+
\sum\limits_{m=1,2}\frac{\left|g_{32}^{k_{2} } \right|^{2} }{\Delta _{m}^{\left(2\right)} } \sum\limits_{i_{m} j_{m} }e^{i\vec{k}_{2} \vec{r}_{i_{m} j_{m} } } S_{32}^{i_{m} } S_{23}^{j_{m} }   +
\sum\limits_{m=1,2}\frac{\left|g_{32}^{k_{2} } \right|^{2} }{\Delta _{m}^{\left(2\right)} } \sum\limits_{j_{m} }\left(S_{33}^{i_{m} } -S_{22}^{i_{m} } \right)n_{k_{2} } \\
+
\frac{1}{2} \left(\frac{1}{\Delta _{1}^{\left(2\right)} } +\frac{1}{\Delta _{2}^{\left(2\right)} } \right)\left|g_{32}^{k_{3} } \right|^{2}
\sum\limits_{j_{1} j_{2} }\left(e^{i\vec{k}_{2} \vec{r}_{j_{1} j_{2} } } S_{32}^{j_{1} } S_{23}^{j_{2} } +h.c.\right)\\
-
\frac{1}{2} \sum\limits_{m=1,2}\frac{1}{\Delta _{m}^{\left(2\right)} }
\sum\limits_{j_{1} }\left(g_{32}^{k_{2} *} g_{31}^{k_{3} } e^{i\left(\vec{k}_{3} -\vec{k}_{2} \right)\vec{r}_{j_{1} } } S_{21}^{i_{1} } a_{k_{3} } a_{k_{2} }^{+} +h.c.\right) \\
-
\frac{1}{2}\sum\limits_{m=1,2}\frac{1}{\Delta _{m}^{\left(2\right)} }  \sum\limits_{j_{2} }\left(g_{32}^{k_{2} } g_{21}^{k_{1} } e^{i\left(\vec{k}_{3} -\vec{k}_{2} \right)\vec{r}_{j_{2} } } S_{31}^{i_{2} } a_{k_{1} } a_{k_{2} } +h.c.\right) \\
+
\sum\limits_{m=1,2}\frac{\left|g_{31}^{k_{3} } \right|^{2} }{\Delta _{m}^{\left(3\right)} } \sum\limits_{i_{m} j_{m} }e^{i\vec{k}_{3} \vec{r}_{i_{m} j_{m} } } S_{31}^{i_{m} } S_{13}^{j_{m} } +
\sum\limits_{m=1,2}\frac{\left|g_{31}^{k_{3} } \right|^{2} }{\Delta _{m}^{\left(3\right)} } \sum\limits_{j_{m} }\left(S_{33}^{j_{m} } -S_{11}^{j_{m} } \right)n_{k_{3} } \\
+
\frac{1}{2} \left(\frac{1}{\Delta _{1}^{\left(3\right)} } +\frac{1}{\Delta _{2}^{\left(3\right)} } \right)\left|g_{31}^{k_{3} } \right|^{2}
\sum\limits_{j_{1} j_{2} }\left(e^{i\vec{k}_{3} \vec{r}_{j_{1} j_{2} } } S_{31}^{j_{1} } S_{13}^{j_{2} } +h.c.\right) \\
-
\frac{1}{2} \sum\limits_{m=1,2}\frac{1}{\Delta _{m}^{\left(3\right)} }
\sum\limits_{j_{1} }\left(g_{32}^{k_{2} } g_{31}^{k_{3} *} e^{i\left(\vec{k}_{2} -\vec{k}_{3} \right)\vec{r}_{j_{1} } } S_{12}^{j_{1} } a_{k_{2} } a_{k_{3} }^{+} +h.c.\right)\\
-
\frac{1}{2} \sum\limits_{m=1,2}\frac{1}{\Delta _{m}^{\left(3\right)} }  \sum\limits_{j_{1} }\left(g_{21}^{k_{1} } g_{31}^{k_{3} *} e^{i\left(\vec{k}_{1} -\vec{k}_{3} \right)\vec{r}_{j_{1} } } S_{23}^{j_{1} } a_{k_{1} } a_{k_{3} }^{+} +
h.c.\right)\, ,
 \end{array}
\end{equation}

where $\Delta _{1,2}^{\left(1\right)} =\varepsilon _{1,2}^{\left(2\right)}
-\varepsilon _{1,2}^{\left(1\right)} -\hbar \omega _{k_{1} } $, $\Delta
_{1,2}^{\left(2\right)} =\varepsilon _{1,2}^{\left(3\right)} -\varepsilon
_{1,2}^{\left(2\right)} -\hbar \omega _{k_{2} } $, $\Delta
_{1,2}^{\left(3\right)} =\varepsilon _{1,2}^{\left(3\right)} -\varepsilon
_{1,2}^{\left(1\right)} -\hbar \omega _{k_{3} } $.

In the absence of real photons in the cavity $n_{k_{1} } =n_{k_{2} } =n_{k_{3}
} =0$, we get from \eqref{Equation-9_} the following expression for effective
ET Hamiltonian:

\begin{equation} \label{Equation-10_} \begin{array}{l} {H_{s} =\sum\limits_{m}^{1,2}\sum\limits_{j_{m} \mu }\varepsilon _{m}^{\mu } S_{\mu \mu }^{j_{m} }   }  +
\sum\limits_{m=1,2}\frac{\left|g_{21}^{k_{3} } \right|^{2} }{\Delta _{m}^{\left(1\right)} } \sum\limits_{i_{m} j_{m} }e^{i\vec{k}_{1} \vec{r}_{i_{m} j_{m} } } S_{21}^{i_{m} } S_{12}^{j_{m} } \\
+
\frac{1}{2} \left(\frac{1}{\Delta _{1}^{\left(1\right)} } +\frac{1}{\Delta _{2}^{\left(1\right)} } \right)\left|g_{21}^{k_{1} } \right|^{2}
\sum\limits_{j_{1} j_{2} }\left(e^{i\vec{k}_{1} \vec{r}_{j_{1} j_{2} } } S_{21}^{j_{1} } S_{12}^{j_{2} } +e^{-i\vec{k}_{1} \vec{r}_{j_{1} j_{2} } } S_{12}^{j_{1} } S_{21}^{j_{2} } \right) \\
+
\sum\limits_{m=1,2}\frac{\left|g_{32}^{k_{2} } \right|^{2} }{\Delta _{m}^{\left(2\right)} } \sum\limits_{i_{m} j_{m} }e^{i\vec{k}_{2} \vec{r}_{i_{m} j_{m} } } S_{32}^{i_{m} } S_{23}^{j_{m} }\\
+
\frac{1}{2} \left(\frac{1}{\Delta _{1}^{\left(2\right)} } +\frac{1}{\Delta _{2}^{\left(2\right)} } \right)\left|g_{32}^{k_{3} } \right|^{2}
\sum\limits_{j_{1} j_{2} }\left(e^{i\vec{k}_{2} \vec{r}_{j_{1} j_{2} } } S_{32}^{j_{1} } S_{23}^{j_{2} } +e^{-i\vec{k}_{2} \vec{r}_{j_{1} j_{2} } } S_{23}^{j_{1} } S_{32}^{j_{2} } \right) \\
+
\sum\limits_{m=1,2}\frac{\left|g_{31}^{k_{3} } \right|^{2} }{\Delta _{m}^{\left(3\right)} } \sum\limits_{i_{m} j_{m} }e^{i\vec{k}_{3} \vec{r}_{i_{m} j_{m} } } S_{31}^{i_{m} } S_{13}^{j_{m} }\\
+
\frac{1}{2} \left(\frac{1}{\Delta _{1}^{\left(3\right)} } +\frac{1}{\Delta _{2}^{\left(3\right)} } \right)\left|g_{31}^{k_{3} } \right|^{2}
\sum\limits_{j_{1} j_{2} }\left(e^{i\vec{k}_{3} \vec{r}_{j_{1} j_{2} } } S_{31}^{j_{1} } S_{13}^{j_{2} } +e^{-i\vec{k}_{3} \vec{r}_{i_{1} j_{1} } } S_{13}^{j_{1} } S_{31}^{j_{2} } \right) \, \, .
\end{array}
\end{equation}

\end{appendix}

\end{document}